\documentclass[a4paper,11pt]{article}
\usepackage{jheppub}
\usepackage{multirow}
\allowdisplaybreaks

\newcommand{\beq}{\begin{equation}}
\newcommand{\eeq}{\end{equation}}
\newcommand{\beqa}{\begin{eqnarray}}
\newcommand{\eeqa}{\end{eqnarray}}

\newcommand{\cm}{{\cal M}^0}

\newcommand{\ds}{{\rm d}\hat{\sigma}}
\newcommand{\dphi}{{\rm d}\Phi}

\newcommand{\re}{{\rm{Re}}}
\newcommand{\norm}{{\cal N}}

\newcommand{\q}[1]{#1_q}

\newcommand{\Q}[1]{#1_Q}
\newcommand{\Qb}[1]{#1_{\bar{Q}}}
\newcommand{\qi}[1]{\hat{#1}_q}
\newcommand{\qbi}[1]{\hat{#1}_{\bar{q}}}

\newcommand{\cep}{C(\epsilon)}
\newcommand{\cepb}{\bar{C}(\epsilon)}
\newcommand{\asmu}{\alpha_s(\mu)}

\def\e{\epsilon}

\newcommand{\LO}{\mathrm{LO}}
\newcommand{\NLO}{\mathrm{NLO}}
\newcommand{\NNLO}{\mathrm{NNLO}}

\newcommand{\Ga}{\Gamma}

\newcommand{\ep}{\epsilon}

\newcommand{\nn}{\nonumber}

\title{Antenna subtraction at NNLO for heavy quark pair production at hadron colliders: the non-diagonal channels}

\preprint{ZU-TH 24/26}

\author{Aude Gehrmann–De Ridder$^{a,b}$, Jo\~{a}o Pires$^{c,d}$}

\affiliation{
$^a$Institute for Theoretical Physics, ETH, CH-8093 Z\"urich, Switzerland\\
        $^b$Physik-Institut, Universit\"at Z\"urich, Winterthurerstrasse 190, 8057 Z\"urich, Switzerland\\
        $^c$LIP, Avenida Professor Gama Pinto 2, P-1649-003 Lisboa, Portugal\\
        $^d$Faculdade de Ciências, Universidade de Lisboa, 1749-016 Lisboa, Portugal
}

\emailAdd{gehra@phys.ethz.ch}
\emailAdd{jnpires@lip.pt}

\abstract{
We present the calculation involving the $qq$, $qq'$
and $qg$ partonic channels contributing to heavy-quark pair production at hadron colliders through next-to-next-to-leading order in QCD using the antenna subtraction formalism.
The calculation is performed in full colour and implemented within the NNLOJET framework. 
Building upon previously available results, we complete the NNLO treatment of the $qq$ and $qq'$ channels by deriving the integrated subtraction contributions required at the real-virtual and double-virtual levels. 
We further present for the first time the NNLO antenna subtraction terms related to the $qg$-channel in full colour. 
The calculation requires several new ingredients, including massive soft factors and convolutions involving integrated massive antennae thereby extending the integrated antenna library to cope with processes with massive fermions. We validate the calculation through analytic infrared pole cancellation and numerical comparisons with existing results. These results complete the NNLO description of the $qq$, $qq'$, and $qg$ channels and constitute an important step towards a fully differential NNLO treatment of heavy-quark pair production within the antenna subtraction formalism.

}


\begin{document} 

\maketitle

\section{Introduction}
\label{sec:intro}    

Precision has become one of the defining features of the physics programme at the Large Hadron Collider, a programme that will be further extended by the High-Luminosity LHC upgrade. Measurements of benchmark Standard Model processes are now routinely performed with uncertainties at the few-percent level, placing increasingly stringent demands on the accuracy of theoretical predictions. Matching this experimental precision is essential both for testing Quantum Chromodynamics (QCD) at the highest energies and for extracting the fundamental parameters of the Standard Model. For many hadron-collider observables, this requires the inclusion of radiative corrections beyond next-to-leading order, making the development of efficient higher-order computational methods an indispensable ingredient of precision phenomenology.

Among these processes, top-quark pair production occupies a central role in the LHC physics programme. Owing to its large production rate and clean experimental signature, it provides one of the most precise probes of the Standard Model. Measurements of inclusive and differential $t\bar{t}$ cross sections are used to determine the top-quark mass, constrain the gluon parton distribution function, and provide complementary information on the strong coupling constant, while simultaneously offering sensitivity to physics beyond the Standard Model. To fully exploit the precision of these measurements, equally precise theoretical predictions are required. A central ingredient of such predictions is the fixed-order perturbative calculation of the hard-scattering process within the framework of perturbative QCD.

The differential hadronic cross section for top-quark pair production at hadron colliders is characterised by the hard scale set by the top-quark mass and can be written in the factorised form,
\begin{equation}
d\sigma(\alpha_s(\mu),\mu,m_t,P_1,P_2)
=
\sum_{i,j}
\int
\frac{d\xi_1}{\xi_1}
\frac{d\xi_2}{\xi_2}
\,f_i(\xi_1,\mu)\,
f_j(\xi_2,\mu)\,
d\hat{\sigma}_{ij}(\alpha_s(\mu),\mu,m_t,p_1,p_2).
\end{equation}

In this equation, $P_1=(\sqrt{s}/2,0,0,\sqrt{s}/2)$ and  
$P_2=(\sqrt{s}/2,0,0,-\sqrt{s}/2)$
are the momenta of the incoming
protons, and $p_1=\xi_1P_1$, $p_2=\xi_2P_2$ denote the momenta of the partons  entering the hard scattering.
The parton distribution function $f_i(\xi_i,\mu_F)$ gives the probability of 
finding a parton of flavour $i$ in the proton carrying a momentum fraction $\xi_i$, while ${\rm d}\hat{\sigma}_{ij}$ denotes the partonic cross sections for partons $i$ and $j$ to produce a
$t\bar{t}$ pair. They depend on the renormalisation and factorisation scales, 
$\mu_R$ and $\mu_F$, which we set equal $\mu_R=\mu_F=\mu$.
Each partonic cross section has the following perturbative expansion in powers of the strong coupling constant,
\begin{equation}
d\hat{\sigma}_{ij}
=
\left(4\pi\alpha_s(\mu)\right)^2
\left(
d\hat{\sigma}_{ij,\mathrm{LO}}
+
\left(\frac{\alpha_s(\mu)}{2\pi}\right)
d\hat{\sigma}_{ij,\mathrm{NLO}}
+
\left(\frac{\alpha_s(\mu)}{2\pi}\right)^2
d\hat{\sigma}_{ij,\mathrm{NNLO}}
+
\mathcal{O}\!\left(\alpha_s^3\right)
\right).
\label{eq:partonic_xsec_expansion}
\end{equation}

A summary of the partonic channels present up to $O(\alpha_s^4)$ for top-antitop pair production at LHC is given in 
Table~\ref{tab:ttbar_channels} as shown in~\cite{abelof2013qcd}.\begin{table}[t]
\centering
\footnotesize
\renewcommand{\arraystretch}{1.25}
\setlength{\tabcolsep}{4pt}

\newcommand{\TallCell}[2]{%
  \begin{tabular}{@{}l@{}}
  \rule{0pt}{3.2ex}#1\\[4pt]
  #2\rule[-1.4ex]{0pt}{0pt}
  \end{tabular}%
}

\begin{tabular}{|c|c|cc|ccc|}
\hline
\multirow{3}{*}{\textbf{Channel}}
&
\multirow{3}{*}{$\mathcal{O}(\alpha_s^2)$}
&
\multicolumn{2}{c|}{$\mathcal{O}(\alpha_s^3)$}
&
\multicolumn{3}{c|}{$\mathcal{O}(\alpha_s^4)$}
\\
\cline{3-7}

&
&
Real
&
Virtual
&
Double real
&
Real--virtual
&
Double virtual
\\
\cline{3-7}

&
&
$\mathcal{M}^{0}_{ij\rightarrow t\bar{t}k}$
&
$\mathcal{M}^{1}_{ij\rightarrow t\bar{t}}$
&
$\mathcal{M}^{0}_{ij\rightarrow t\bar{t}kl}$
&
$\mathcal{M}^{1}_{ij\rightarrow t\bar{t}k}$
&
$\mathcal{M}^{2}_{ij\rightarrow t\bar{t}}$
\\
\hline

$q\bar q$
&
$q\bar q\rightarrow t\bar t$
&
$q\bar q\rightarrow t\bar t g$
&
$q\bar q\rightarrow t\bar t$
&
\TallCell
{$q\bar q\rightarrow t\bar t gg$}
{$q\bar q\rightarrow t\bar t q'\bar q'$}
&
$q\bar q\rightarrow t\bar t g$
&
$q\bar q\rightarrow t\bar t$
\\
\hline

$gg$
&
$gg\rightarrow t\bar t$
&
$gg\rightarrow t\bar t g$
&
$gg\rightarrow t\bar t$
&
\TallCell
{$gg\rightarrow t\bar t gg$}
{$gg\rightarrow t\bar t q\bar q$}
&
$gg\rightarrow t\bar t g$
&
$gg\rightarrow t\bar t$
\\
\hline

$qg~(\bar q g)$
&
---
&
$qg\rightarrow t\bar t q$
&
---
&
$qg\rightarrow t\bar t q g$
&
$qg\rightarrow t\bar t q$
&
---
\\
\hline

$qq'~(q\bar q',\,\bar q\bar q')$
&
---
&
---
&
---
&
$qq'\rightarrow t\bar t q q'$
&
---
&
---

\\
\hline

\end{tabular}

\caption{Summary of the different partonic channels and matrix elements that contribute to $t\bar t$ hadronic production through $\mathcal{O}(\alpha_s^4)$.}
\label{tab:ttbar_channels}
\end{table}
In this paper, we shall focus on top-pair production associated with the quark-gluon and quark-quark channels, focussing on the contributions arising from the last two rows in Table~\ref{tab:ttbar_channels}.  The next-to-leading order (${\cal O}(\alpha_s^3)$) contributions are discussed in Section~\ref{sec:nlo}, while the next-to-next-to-leading order (${\cal O}(\alpha_s^4)$) corrections are presented in Section~\ref{sec:ttbnnlo}.

Beyond leading order, the computation of the cross sections for these partonic channels requires the treatment of ultraviolet and infrared singularities. Ultraviolet poles are removed by renormalisation, while the infrared singularities are cancelled through the combination of virtual and real-radiation contributions together with mass factorisation counterterms, in accordance with the Kinoshita–Lee–Nauenberg 
theorem~\cite{Kinoshita:1962ur,Lee:1964is}.

While infrared singularities from purely virtual corrections become explicit immediately after integration over the loop momenta, their extraction is more involved for real-emission and mixed real-virtual contributions. In these cases, the singular behaviour only becomes manifest after integrating the corresponding matrix elements over the phase space relevant to the differential observable under consideration. Since such observables generally depend in a non-trivial way on the experimental selection criteria and kinematic cuts used to define them, these phase-space integrations cannot, in general, be performed analytically and must instead be evaluated numerically. Differential predictions at higher orders in perturbation theory are therefore obtained using a parton-level event generator, which performs the phase-space integration by means of Monte Carlo techniques while implementing the desired experimental cuts and allowing arbitrary infrared-safe observables to be evaluated.

A variety of methods have been developed in the past to 
systematically handle infrared singularities in fully differential calculations computed with parton-level event generators.
At NNLO, local subtraction approaches rely on the derivation of counterterms that reproduce the singular behaviour of the real-radiation matrix elements in every unresolved limit while remaining sufficiently simple to be integrated analytically over the unresolved phase space. Moreover, these counterterms must capture only the singular behaviour they are designed to subtract, without introducing spurious infrared singularities in other regions of phase space.

Those subtraction methods 
include antenna
subtraction~\cite{Gehrmann-DeRidder:2005btv,Currie:2013vh},
sector-improved residue subtraction~\cite{Czakon:2011ve,Czakon:2014oma},
nested soft-collinear subtraction~\cite{Caola:2017dug,Devoto:2025jql},
local analytic sector subtraction~\cite{Magnea:2020trj},
and colour-full subtraction~\cite{DelDuca:2025yph}. 
Other NNLO methods include the projection-to-Born approach~\cite{Cacciari:2015jma} and phase-space slicing methods based on resolution variables, such as $q_T$ subtraction~\cite{Catani:2007vq} and $N$-jettiness subtraction~\cite{Boughezal:2015dva,Gaunt:2015pea}.

In the present paper, we employ the antenna subtraction formalism for processes involving the production of a heavy-quark pair at hadron colliders. The extension of antenna subtraction to massive fermions has been developed progressively through the derivation of massive antenna functions, their integrated counterparts, phase-space factorisations and momentum mappings, as well as through applications to individual contributions to heavy-quark production at hadron colliders~\cite{Gehrmann-DeRidder:2009lyc,Abelof:2011ap,Abelof:2012rv,Abelof:2012he,Abelof:2014fza,Abelof:2014jna,Abelof:2015lna}. The same formalism has also been employed in a fully differential NNLO QCD calculation of $e^+e^- \to t\bar{t}$~\cite{Chen:2016zbz,Dekkers:2014hna,Bernreuther:2013uma}. In this paper, we further advance the massive antenna subtraction formalism by deriving the building blocks required for the NNLO treatment of the flavour off-diagonal channels in heavy-quark pair production.

NNLO subtraction methods for processes involving massive particles follow the same general strategy as in the massless case, but the pattern of infrared singularities is modified by the presence of finite masses and the infrared structure of the corresponding amplitudes is qualitatively different. While QCD radiation from massive particles can still give rise to soft singularities, the heavy-quark mass strictly regulates the collinear singularities. Collinear configurations involving massive partons are replaced by quasi-collinear configurations, which are integrable but give rise to finite logarithmic contributions involving ratios of the heavy-quark mass to the other kinematic scales of the process~\cite{Catani:2002hc,Catani:2000ef}. These finite logarithmic contributions are the remnants of the collinear singularities that would be present in the massless limit and become enhanced only in kinematic regions where widely separated scales are involved. In the kinematical regime considered in this paper, namely the production of a heavy-quark pair with a mass of the same order as the partonic centre-of-mass energy, such logarithmic enhancements are not expected to play a significant role and will therefore be neglected. The same approach has been adopted in all calculations of top-antitop pair production~\cite{Czakon:2015owf,Catani:2019hip}. Consequently, subtraction terms are required only to capture soft singularities between massless or massive partons and to deal with collinear radiation involving massless partons.

At NNLO, the partonic contributions to a hadronic observable
whose Born-level final state consists of a heavy-quark pair together with $(m-2)$ massless partons can be written as,
\beqa
{\rm d}\hat\sigma_{NNLO}&=&\int_{{\rm{d}}\Phi_{m+2}} {\rm{d}}\hat\sigma_{NNLO}^{RR}
+\int_{{\rm{d}}\Phi_{m+1}} {\rm{d}}\hat\sigma_{NNLO}^{RV} 
+\int_{{\rm{d}}\Phi_m}{\rm{d}}\hat\sigma_{NNLO}^{VV},
\label{eq:subnnlo1}
\eeqa
where the three terms correspond to the double-real, mixed real-virtual and
double-virtual contributions, ${\rm{d}}\hat\sigma_{NNLO}^{RR}$, ${\rm{d}}\hat\sigma_{NNLO}^{RV}$ and ${\rm{d}}\hat\sigma_{NNLO}^{VV}$, 
respectively. Each of these contributions is separately infrared divergent. 

Employing a subtraction method, eq.(\ref{eq:subnnlo1}) 
amounts to~\cite{Gehrmann-DeRidder:2005btv}
\beqa
{\rm d}\hat\sigma_{NNLO}&=&\int_{{\rm{d}}\Phi_{m+2}}\left({\rm{d}}\hat\sigma_{NNLO}^{RR}-{\rm{d}}\hat\sigma_{NNLO}^S\right)
+\int_{{\rm{d}}\Phi_{m+2}}{\rm{d}}\hat\sigma_{NNLO}^S\nonumber\\
&+&\int_{{\rm{d}}\Phi_{m+1}}\left({\rm{d}}\hat\sigma_{NNLO}^{RV}-{\rm{d}}\hat\sigma_{NNLO}^{V S}\right)
+\int_{{\rm{d}}\Phi_{m+1}}{\rm{d}}\hat\sigma_{NNLO}^{V S}
+\int_{{\rm{d}}\Phi_{m+1}}{\rm{d}}\hat\sigma_{NNLO}^{MF,1}\nonumber\\
&+&\int_{{\rm{d}}\Phi_m}{\rm{d}}\hat\sigma_{NNLO}^{VV}
+\int_{{\rm{d}}\Phi_m}{\rm{d}}\hat\sigma_{NNLO}^{MF,2}\label{eq.sigNNLO},
\eeqa
where ${\rm d} \hat\sigma^{S}_{NNLO}$ denotes the subtraction term for the $(m+2)$-parton final state which behaves like the double real radiation contribution ${\rm d} \hat\sigma^{RR}_{NNLO}$ in all singular limits. Likewise, ${\rm d} \hat\sigma^{VS}_{NNLO}$ is the one-loop virtual subtraction term coinciding with the one-loop $(m+1)$-final state ${\rm d} \hat\sigma^{RV}_{NNLO}$ in all singular limits. The two-loop correction to the $(m+2)$-parton final state is denoted by ${\rm d}\hat\sigma^{VV}_{NNLO}$.  In addition, as there are partons in the initial state, there are also two mass factorisation contributions, ${\rm d}\hat\sigma^{MF,1}_{NNLO}$ and ${\rm d}\hat\sigma^{MF,2}_{NNLO}$, for the $(m+1)$- and $m$-particle final states respectively.

In order to compute the NNLO cross section, the various contributions must be reorganised according to the number of final state particles as explained in detail for di-jet observables in~\cite{NigelGlover:2010kwr,Gehrmann-DeRidder:2011jwo,Gehrmann-DeRidder:2012dog,Currie:2013dwa}, 
and in heavy-quark pair production
in~\cite{Abelof:2014fza,Abelof:2015lna}. 
The NNLO partonic cross section for top-pair production in a given partonic channel
(and proportional to a specific colour factor) reads, 
\beqa
\ds_{NNLO}&=&\int_{{\rm{d}}\Phi_{m+2}}\left[\ds_{NNLO}^{RR}-\ds_{NNLO}^S\right]
\nonumber \\
&+& \int_{{\rm{d}}\Phi_{m+1}}
\left[
\ds_{NNLO}^{RV}-\ds_{NNLO}^{T}
\right] \nonumber \\
&+&\int_{{\rm{d}}\Phi_{m\phantom{+1}}}\left[
\ds_{NNLO}^{VV}-\ds_{NNLO}^{U}\right],\label{eq.subnnlo}
\eeqa
where the terms in each of the square brackets are finite and well behaved in the infrared singular regions. The general structure of the NNLO subtraction terms
${\rm d}\hat{\sigma}_{NNLO}^{S}$,
${\rm d}\hat{\sigma}_{NNLO}^{T}$ and ${\rm d}\hat{\sigma}_{NNLO}^{U}$, and their organisation within the antenna-subtraction formalism are discussed in detail in Refs.~\cite{Gehrmann-DeRidder:2005btv,Currie:2013vh} and will not be repeated here.

For processes involving massive fermions, the antenna subtraction formalism was established in full generality at NLO in Ref.~\cite{Gehrmann-DeRidder:2009lyc,Abelof:2011jv}. Subsequent work extended the formalism to NNLO, providing the subtraction terms and integrated ingredients for selected partonic channels and colour structures relevant to heavy-quark pair production at hadron colliders~\cite{Abelof:2011ap,Abelof:2012rv,Abelof:2012he,Abelof:2014fza,Abelof:2014jna,Abelof:2015lna}. 

The aim of this paper is to complete the antenna subtraction treatment of  the $qq'$, $qq$ and $qg$ partonic channels contributing to heavy-quark pair production up to ${\cal O}(\alpha_s^4)$ in QCD. A key new feature of the present calculation is the appearance of new massive soft factors and their integrated counterparts, which were not required in previous NNLO antenna-subtraction calculations for top-pair production.

The remainder of this paper is organised as follows. In Section~\ref{sec:nlo}, we introduce the notation and conventions used throughout the paper and review the infrared structure of top-pair production in the $qg$ channel at NLO, thereby establishing the framework on which the subsequent NNLO calculation is based. Section~\ref{sec:ttbnnlo} presents the NNLO antenna-subtraction 
treatment of the $qq'$, $qq$ and $qg$ partonic channels, including the corresponding subtraction terms and their infrared structure. In Section~\ref{sec:massiveSF}, we derive the new massive soft factors and their integrated counterparts required for the NNLO antenna-subtraction treatment of these channels, and present their numerical validation. Section~\ref{sec:nnlonumerics} presents the numerical results obtained with NNLOJET and compares them with independent predictions from 
{\tt Top++}~\cite{Czakon:2011xx}. Finally, our conclusions are given in Section~\ref{sec:conclusions}.
\section {Infrared structure of top-pair production in the $qg$-channel up to ${\cal O}(\alpha_s^3$) }
\label{sec:nlo}

\subsection{Top-pair production at ${\cal O}(\alpha_s^2)$}
Let us begin by considering the cross section for top-pair production at leading order at hadron-colliders, from which the normalization conventions and Born matrix elements entering the subtraction terms at higher orders are defined.
\newline
The hadronic cross section for $t\bar{t}$ production at leading order involves two partonic channels, with either a $q\bar{q}$ pair or a pair of gluons in the initial state. It is given by  
\beqa
&&\hspace{-0.5in}{\rm d}\sigma_{\LO}(H_1,H_2)=\int\frac{{\rm d}\xi_1}{\xi_1}\frac{{\rm d}\xi_2}{\xi_2}\bigg( f_g(\xi_1,\mu)f_g(\xi_2,\mu)\:\ds_{gg,\LO}(p_1,p_2)\nonumber\\
&&\hspace{0.965in}+\sum_q f_q(\xi_1,\mu)f_{\bar{q}}(\xi_2,\mu)\:\ds_{q\bar{q},\LO}(p_1,p_2) \bigg),
\eeqa
where $H_1$ and $H_2$ are the momenta of the incoming hadrons, $p_i=\xi_i H_i$, $f_i(\xi,\mu)$ are the parton distribution functions evaluated at the factorisation scale $\mu$,
and the sum runs over all light quark flavours. 

\subsubsection{The $q\bar{q}$ channel}
The leading order (LO) partonic cross section for the $q \bar{q}$ initiated process takes the form: 
\beq\label{eq.qqblo}
\ds_{q\bar{q},\LO}=\norm_{\LO}^{\:q\bar{q}}\int\dphi_2(p_3,p_4;p_1,p_2)\:|\cm_4(\Q{3},\Qb{4},\qbi{2},\qi{1})|^2 J^{(2)}_2(p_3,p_4),
\eeq
where, $\dphi_2(p_3,p_4;p_1,p_2)$ is the $2 \to 2$ partonic phase space, $J^{(2)}_2(p_3,p_4)$ is a so-called measurement function, which ensures that a pair of final state massive quarks of momenta $p_3$ and $p_4$ are observed. Here and throughout the paper, $Q(\bar{Q})$ denote a heavy (anti)quark, and initial-state partons are indicated by a hat on the corresponding momentum labels, e.g., $\hat{1}_q,\hat{2}_{\bar{q}}$.  $\cm_4(...)$ is the colour-ordered and coupling-stripped tree-level amplitude.  It is related to the full amplitude through the (trivial) colour decomposition
\beq\label{eq.coldecqqblo}
M_4^0(q_1\bar{q}_2\rightarrow Q_3 \bar{Q}_4)=g_s^2\left( \delta_{i_3i_1}\delta_{i_2i_4}-\frac{1}{N_c}\delta_{i_3i_4}\delta_{i_2i_1}\right)\cm_4(\Q{3},\Qb{4},\qbi{2},\qi{1}).
\eeq
The normalisation factor is
\beq\label{eq.normlo}
\norm_{\LO}^{\:q\bar{q}}=\frac{1}{2s}\:\left( \frac{\asmu}{2\pi}\right)^2\:\frac{\cepb^2}{\cep^2}\:\frac{(N_c^2-1)}{4N_c^2},
\eeq
where $s$ denotes the hadronic center-of-mass energy squared, and $N_c$ the number of colors, with $N_c=3$ in QCD. Included in this normalisation factor are the flux factor, as well as the sum and average over colour and spin.

The constants $\cep$ and $\cepb$  are defined as:
\beq
\cep=\frac{(4\pi)^{\e}}{8\pi^2}e^{-\e \gamma_E} \hspace{1.5in}   
\cepb=(4\pi)^{\e} e^{-\e \gamma_E},
\eeq
providing the useful relation
\beq
g_s^2=4\pi\alpha_s=\left( \frac{\alpha_s}{2\pi}\right)\frac{\cepb}{\cep}.
\eeq

Using the NNLOJET notation~\cite{NNLOJET:2025rno}, the coupling and colour-stripped matrix-element squared contribution 
associated to this four quark process is of $C$-type and reads:  
\begin{eqnarray}
|\cm_4(\Q{3},\Qb{4},\qbi{2},\qi{1})|^2&=& C_{0}^0(\hat{1}_q,3_Q,4_{\bar{Q}},\hat{2}_{\bar{q}})\,,
\end{eqnarray} 
where the superscript labels the loop order, while the subscript denotes the number of external gluons accompanying the two quark-antiquark pairs.

\subsubsection{The $gg$ channel}
The colour decomposition of the $2 \to 2$ tree level amplitude for the partonic process $gg \to t \bar{t}$ is given by,

\beq\label{eq.coldecgglo}
M_4^0(g_1g_2\rightarrow Q_3 \bar{Q}_4)=g_s^2\,2\sum_{\sigma_2}
(T^{a_{\sigma(1)}}T^{a_{\sigma(2)}})_{i_3i_4}
\cm_4(\Q{3},\sigma(1)_g,\sigma(2)_g,\Qb{4})\,
\eeq
where $\sigma_2$ is the set of permutations of 2-gluons
and $T^a$ the generators of the fundamental representation of
$SU(N_c)$.
In terms of colour-ordered amplitudes the LO partonic cross section
in the $gg$-channel can be written as,
\beqa
\label{eq.ggblo}
\ds_{gg,\LO}=\norm_{\LO}^{\:gg}\int\dphi_2(p_3,p_4;p_1,p_2)
&&\Big\{N_c\;\Big( B_2^0(3_Q,\hat{1}_g,\hat{2}_g,4_{\bar{Q}})
+B_2^0(3_Q,\hat{2}_g,\hat{1}_g,4_{\bar{Q}})\Big)\nonumber\\
&&-\frac{1}{N_c}
\tilde{B}_2^0(3_Q,\hat{1}_\gamma,\hat{2}_\gamma,4_{\bar{Q}})
\Big\}J^{(2)}_2(p_3,p_4)\,.
\eeqa
Using the NNLOJET notation~\cite{NNLOJET:2025rno}, squared
matrix elements containing a single quark-antiquark pair accompanied by an arbitrary number of gluons are classified as $B$-type with 
subleading-colour contributions indicated by a tilde, e.g. $\tilde{B}$.
The leading colour contribution is therefore defined as, 
\beqa
B_2^0(3_Q,\hat{1}_g,\hat{2}_g,4_{\bar{Q}})=
|\cm_4(\Q{3},\hat{1}_g,\hat{2}_g,\Qb{4})|^2\;
\eeqa
and in the sub-leading-colour contribution,
\beqa
\cm_4(\Q{3},\hat{1}_\gamma,\hat{2}_\gamma,\Qb{4})&=&
\cm_4(\Q{3},\hat{1}_g,\hat{2}_g,\Qb{4})
+\cm_4(\Q{3},\hat{2}_g,\hat{1}_g,\Qb{4}),\nonumber\\
\tilde{B}_2^0(3_Q,\hat{1}_\gamma,\hat{2}_\gamma,4_{\bar{Q}})&=&
|\cm_4(\Q{3},\hat{1}_\gamma,\hat{2}_\gamma,\Qb{4})|^2,
\eeqa
and both gluons are $U(1)$-like and do not couple to each other.
Notationwise, we denote gluons which are photon-like and only
couple to quark lines, with the index $\gamma$  instead of $g$, to manifestly separate leading from 
subleading contributions. The normalisation factor in this channel is

\beq\label{eq.normlogg}
\norm_{\LO}^{\:gg}=\frac{1}{2s}\:\left( \frac{\asmu}{2\pi}\right)^2\:\frac{\cepb^2}{\cep^2}\:\frac{(N_c^2-1)}{4(N_c^2-1)^2},
\eeq
where the only difference with respect to the $q\bar{q}$-channel is in
the colour averaging, which now yields a factor of $(N_c^2-1)^2$ in the denominator.

\subsection{Top-pair production at ${\cal O}(\alpha_s^3)$ }
At the next-to-leading order, the $q\bar{q}$ and $gg$ channels receive 
virtual and real corrections, while a new quark-gluon initiated 
channel contributes for the first time, see Table 
\ref{tab:ttbar_channels}. By the $qg$-channel we 
shall collectively refer to the  partonic initial 
states $qg$, $\bar{q}g$, where the
$\bar{q}g$ contributions are obtained from the $qg$ ones by
charge conjugation. The corresponding $gq$ and $g\bar{q}$ contributions follow by exchanging the incoming parton momenta and will not be discussed separately.
The hadronic cross section for $t\bar{t}$ production at this order is therefore given by,
\beqa
&&\hspace{-0.25in}{\rm d}\sigma_{\NLO}(P_1,P_2)=\int\frac{{\rm d}\xi_1}{\xi_1}\frac{{\rm d}\xi_2}{\xi_2}\bigg[ f_g(\xi_1,\mu)f_g(\xi_2,\mu)\:\ds_{gg,\NLO}(p_1,p_2)\phantom{\sum_q}\nonumber\\
&&\hspace{-0.25in}\phantom{{\rm d}\sigma_{\NLO}(P_1,P_2)}+\sum_q\bigg( f_q(\xi_1,\mu)f_{\bar{q}}(\xi_2,\mu)\:\ds_{q\bar{q},\NLO}(p_1,p_2)\nonumber\\
&&\hspace{-0.25in}\phantom{{\rm d}\sigma_{\NLO}(P_1,P_2)}+ \Big(f_q(\xi_1,\mu)+ f_{\bar{q}}(\xi_1,\mu)\Big)f_{g}(\xi_2,\mu)\ds_{qg,\NLO}(p_1,p_2)
\nonumber\\
&&\hspace{-0.25in}\phantom{{\rm d}\sigma_{\NLO}(P_1,P_2)}+ f_{g}(\xi_1,\mu)\Big(f_q(\xi_2,\mu)+ f_{\bar{q}}(\xi_2,\mu)\Big)\ds_{gq,\NLO}(p_1,p_2)
\bigg)\bigg].
\eeqa
In the antenna subtraction formalism, a generic NLO partonic contribution is written 
as~\cite{Gehrmann-DeRidder:2005btv,Daleo:2006xa,Currie:2013vh},
\beq
{\rm d}\hat{\sigma}_{\NLO}
=
\int_{{\rm d}\Phi_{m+1}}
\left(
{\rm d}\hat{\sigma}^{R}_{\NLO}
-
{\rm d}\hat{\sigma}^{S}_{\NLO}
\right)
+
\int_{{\rm d}\Phi_m}
\left(
{\rm d}\hat{\sigma}^{V}_{\NLO}
+
\int_1 {\rm d}\hat{\sigma}^{S}_{\NLO}
+
{\rm d}\hat{\sigma}^{MF}_{\NLO}
\right).
\label{eq:nlogeneric}
\eeq
The subtraction term ${\rm d}\hat{\sigma}^{S}_{\NLO}$ reproduces the single-unresolved limits of the real-radiation matrix element, such that the first bracket is finite in four dimensions. After integration over the unresolved phase space, $\int_1{\rm d}\hat{\sigma}^{S}_{\NLO}$ contains explicit infrared poles which cancel against those of the virtual contribution and the mass-factorisation counterterm in the second bracket. It is therefore convenient to combine the integrated subtraction term and the mass-factorisation counterterm into the contribution
\beq
{\rm d}\hat{\sigma}^{T}_{\NLO}
=
\int_1{\rm d}\hat{\sigma}^{S}_{\NLO}
+
{\rm d}\hat{\sigma}^{MF}_{\NLO},
\eeq
which contains the integrated subtraction terms required to cancel the explicit infrared poles of the virtual matrix elements.

The general structure outlined above can be applied to the NLO $qg$-channel contribution to heavy-quark pair production in hadronic collisions. We first present the colour decomposition of the real-emission matrix elements, followed by the derivation of the corresponding antenna subtraction terms.

\subsubsection{Real contribution in the $qg$ channel}
As shown in Table~\ref{tab:ttbar_channels} in the introduction, the real radiation corrections for the $qg$-channel are due to the
process $qg\to t\bar{t}q$. The colour decomposition of the 
corresponding tree-level amplitude is,
\beqa
&&M_5^0(q_1g_2\rightarrow Q_3 \bar{Q}_4 q_5)=g_s^3
\sqrt{2}\:\Big\{ \nonumber\\
&&\hspace{0.6cm}N_c\Big[
\left(T^{a_2}\right)_{i_3i_1}\delta_{i_5i_4}
{\cal M}_5^{0,a}(3_Q,\hat{2}_g,\hat{1}_{\bar{q}};;5_q,4_{\bar{Q}})
+\left(T^{a_2}\right)_{i_5i_4}\delta_{i_3i_1}
{\cal M}_5^{0,b}(3_Q,\hat{1}_{\bar{q}};;5_q,\hat{2}_g,4_{\bar{Q}})\Big]\nonumber\\
&&\hspace{0.4cm}-\frac{1}{N_c}
\left(T^{a_2}\right)_{i_3i_4}\delta_{i_5i_1}
{\cal M}_5^{0,c}(3_Q,\hat{2}_g,4_{\bar{Q}};;5_q,\hat{1}_{\bar{q}})
+\left(T^{a_2}\right)_{i_5i_1}\delta_{i_3i_4}
{\cal M}_5^{0,d}(3_Q,4_{\bar{Q}};;5_q,\hat{2}_g,1_{\bar{q}})\Big]\Big\}.\nonumber\\
\label{eq:nloRqg}
\eeqa
To make the colour structure in Eq.~\eqref{eq:nloRqg} explicit, the partons in the colour-ordered amplitudes are separated by a double semicolon, which distinguishes independent colour strings. Partons belonging to the same colour string are colour connected, whereas partons separated by the double semicolon are not. This notation, introduced in Ref.~\cite{Abelof:2015lna}, makes the colour connections underlying the antenna subtraction construction manifest and will be used throughout this paper.

Squaring this expression and combining it with the $2\to3$ 
phase space, the appropriate overall factors and the measurement function, we can write the real radiation contribution as,
\beqa
&&{\rm d}\hat{\sigma}_{qg,NLO}^R=\nonumber\\
&&\hspace{0.6cm}\norm_{\NLO}^{\:qg}\int{\rm d}\Phi_3(p_3,p_4,p_5;p_1,p_2)
\Big\{N_c\Big[
C_{1}^{0,a}(3_Q,\hat{2}_g,\hat{1}_{\bar{q}};;5_q,4_{\bar{Q}})+C_{1}^{0,b}(3_Q,\hat{1}_{\bar{q}};;5_q,\hat{2}_g,4_{\bar{Q}})\Big]\nonumber\\
&&\hspace{1.6cm}+\frac{1}{N_c}\Big[
C_{1}^{0,c}(3_Q,\hat{2}_g,4_{\bar{Q}};;5_q,\hat{1}_{\bar{q}})+C_{1}^{0,d}(3_Q,4_{\bar{Q}};;5_q,\hat{2}_g,\hat{1}_{\bar{q}})\nonumber\\
&&\hspace{2.6cm}
-2C_{1}^{0,\gamma}(3_Q,4_{\bar{Q}},5_q,\hat{1}_{\bar{q}},\hat{2}_\gamma)
\Big]\Big\}J^{(3)}_2(p_3,p_4,p_5)\,,
\label{eq:nloRqg2}
\eeqa
with the colour-stripped matrix elements appearing in Eq.~\eqref{eq:nloRqg2} defined as the squared moduli of the corresponding colour-ordered
amplitudes,
\beq
C_1^{0,x}(\ldots)=\left|{\cal M}_5^{0,x}(\ldots)\right|^2,
\qquad x=a,b,c,d,\gamma
\eeq
and the decoupled amplitude is used,
\beqa
{\cal M}_5^{0,\gamma}(3_Q,4_{\bar{Q}},5_q,\hat{1}_{\bar{q}},\hat{2}_\gamma)&=&{\cal M}_5^{0,a}(3_Q,\hat{2}_g,\hat{1}_{\bar{q}};;5_q,4_{\bar{Q}})
+{\cal M}_5^{0,b}(3_Q,\hat{1}_{\bar{q}};;5_q,\hat{2}_g,4_{\bar{Q}})\nonumber\\
&=&{\cal M}_5^{0,c}(3_Q,\hat{2}_g,4_{\bar{Q}};;5_q,\hat{1}_{\bar{q}})+
{\cal M}_5^{0,d}(3_Q,\hat{2}_g,4_{\bar{Q}};;5_q,\hat{1}_{\bar{q}}),
\eeqa
in which the gluon is $U(1)$-like and only has photon-like couplings. 
In~\eqref{eq:nloRqg2}, $J_2^{(3)}(p_3,p_4,p_5)$ is a 
measurement-function, that
ensures that heavy quarks of momenta $p_3$ and $p_4$ are observed, while the additional massless parton with momentum $p_5$ is treated inclusively and may either be resolved as an extra jet or remain unresolved.
The normalisation factor $\norm_{\NLO}^{\:qg}$ is given by,
\beq
\norm_{\NLO}^{\:qg}=\frac{\norm_{\LO}^{\:q\bar{q}}\,N_c^2}{N_c(N_c^2-1)}
\left(\frac{\alpha_s}{2\pi}\right)\frac{\cepb}{\cep}\,.
\eeq

\subsubsection{NLO antenna subtraction terms for the $qg$-channel}
At NLO, the colour-ordered matrix elements in Eq.~\eqref{eq:nloRqg2}
develop collinear singularities only when the massless final-state quark becomes collinear to either of the initial-state massless partons. 

As there are no soft singularities, one can obtain the subtraction term for the $qg$-channel entirely with massless antennae using massless radiators. We obtain
\beqa
&&\hspace{-0.4cm}{\rm d}\hat{\sigma}^S_{qg,NLO}=\nonumber\\
&&\hspace{0.0cm}-\norm_{\NLO}^{\:qg}\Phi_3(p_3,p_4,p_5;p_1,p_2)
\Big\{N_c\Big[G_3^0(\hat{2}_g,\hat{1}_{\bar{q}},5_q)B_{2}^0(\tilde{3}_Q,\hat{\bar{1}}_g,\hat{\bar{2}}_g,\tilde{4}_{\bar{Q}})
J^{(2)}_2(\tilde{p}_3,\tilde{p}_4)\nonumber\\
&&\hspace{4.8cm}+G_3^0(\hat{2}_g,\hat{1}_{\bar{q}},5_q)B_{2,}^0(\tilde{3}_Q,\hat{\bar{2}}_g,\hat{\bar{1}}_g,\tilde{4}_{\bar{Q}})
J^{(2)}_2(\tilde{p}_3,\tilde{p}_4)\nonumber\\
&&\hspace{4.8cm}+A_3^0(5_q,\hat{2}_{g},\hat{1}_{\bar{q}})
C_{0}^0(\hat{1}_{\bar{q}},\tilde{3}_Q,\tilde{4}_{\bar{Q}},\hat{\bar{2}}_q)J^{(2)}_2(\tilde{p}_3,\tilde{p}_4)
\Big]\nonumber\\
&&\hspace{4.1cm}-\frac{1}{N_c}\Big[
A_3^0(5_q,\hat{2}_g,\hat{1}_{\bar{q}})C_{0}^0(\hat{\bar{1}}_{\bar{q}},\tilde{3}_Q,\tilde{4}_{\bar{Q}},\hat{\bar{2}}_q)
J^{(2)}_2(\tilde{p}_3,\tilde{p}_4)\nonumber\\
&&\hspace{4.8cm}+\frac{1}{2}G_3^0(\hat{2}_g,\hat{1}_{\bar{q}},5_q)\tilde{B}_{2}^0(\tilde{3}_Q,\hat{\bar{1}}_g,\hat{\bar{2}}_g,\tilde{4}_{\bar{Q}})
J^{(2)}_2(\tilde{p}_3,\tilde{p}_4)\nonumber\\
&&\hspace{4.8cm}+\frac{1}{2}G_3^0(\hat{2}_g,\hat{1}_{\bar{q}},5_q)\tilde{B}_{2}^0(\tilde{3}_Q,\hat{\bar{2}}_g,\hat{\bar{1}}_g,\tilde{4}_{\bar{Q}})
J^{(2)}_2(\tilde{p}_3,\tilde{p}_4)\Big]\Big\}.
\label{eq:snloqg}
\eeqa

The subtraction term naturally inherits the colour structure of the real-radiation contribution in Eq.~\eqref{eq:nloRqg2}. The required collinear limits are described by the massless initial-initial $A_3^0$ and $G_3^0$ antenna functions~\cite{Daleo:2006xa} to subtract the $q\to qg$ and $g\to q\bar{q}$
splittings respectively. 

Both singular limits are identity-changing in the initial state, in the sense that clustering the unresolved parton with the initial-state radiator changes its flavour.

After integration over the unresolved phase space, the subtraction term develops explicit infrared poles. In the $qg$-channel, however, there is no one-loop matrix element at this order, so these poles are not cancelled by a virtual contribution. Instead, the initial-state
collinear singularities are removed through mass factorisation and absorbed into the parton distribution functions. 

Following the standard antenna-subtraction formalism, we combine the integrated antenna functions~\cite{Daleo:2006xa}
with the first-order mass-factorisation splitting kernels
$\Gamma_{ij}^{(1)}$, describing the subtraction of initial-state collinear singularities, into the integrated dipole operator
$\mathcal{J}_2^{(1)}$~\cite{Currie:2013vh,Chen:2022clm}.
This organisation makes the infrared singularity structure explicit and provides a compact representation of the integrated subtraction terms.

The real radiation subtraction term~\eqref{eq:snloqg}, integrated over
the single unresolved phase space and combined with the NLO mass-factorisation contribution, yields
\beqa
&&{\rm d}\hat{\sigma}_{NLO,qg}^{T}=
-\norm_{\NLO}^{\:qg}C(\epsilon)\:
{\rm d}\Phi_2(p_3,p_4;p_1,p_2)
\Big\{\nonumber\\
&&\hspace{3.0cm}N_c\Big[
\mathcal{J}_{2,g\to q}^{(1)}(\hat{1}_{q},\hat{2}_{g\to q})
C_0^0(1_q,3_Q,4_{\bar{Q}},2_{\bar{q}})\nonumber\\
&&\hspace{3.3cm}+
\mathcal{J}_{2,q\to g}^{(1)}(\hat{1}_{q\to g},\hat{2}_{g})
B_2^0(3_Q,\hat{1}_g,\hat{2}_g,4_{\bar{Q}})
\nonumber\\
&&\hspace{3.3cm}+
\mathcal{J}_{2,q\to g}^{(1)}(\hat{1}_{q\to g},\hat{2}_{g})
B_2^0(3_Q,\hat{2}_g,\hat{1}_g,4_{\bar{Q}})
\Big]\nonumber\\
&&\hspace{3.0cm}-\frac{1}{N_c}\Big[
\mathcal{J}_{2,g\to q}^{(1)}(\hat{1}_{q},\hat{2}_{g\to q})
C_0^0(1_q,3_Q,4_{\bar{Q}},2_{\bar{q}})\nonumber\\
&&\hspace{3.3cm}+\frac{1}{2}
\mathcal{J}_{2,q\to g}^{(1)}(\hat{1}_{q\to g},\hat{2}_{g})
\tilde{B}_2^0(3_Q,\hat{1}_g,\hat{2}_g,4_{\bar{Q}})\nonumber\\
&&\hspace{3.3cm}+\frac{1}{2}
\mathcal{J}_{2,q\to g}^{(1)}(\hat{1}_{q\to g},\hat{2}_{g})\tilde{B}_2^0(3_Q,\hat{2}_g,\hat{1}_g,4_{\bar{Q}})
\Big]\Big\},
\label{eq:tnloqg}
\eeqa
In this equation, the initial-initial identity changing dipoles are given 
by~\cite{Chen:2022clm},
\beqa
\mathcal{J}_{2,g\to q}^{(1)}(\hat{1}_{q},\hat{2}_{g\to q})
&=&
-{\cal A}_{3,qg}^{0}(s_{12})
-S_{g\to q}\Gamma^{(1)}_{qg}(x_2)
\nonumber\\
\mathcal{J}_{2,q\to g}^{(1)}(\hat{1}_{q\to g},\hat{2}_{g})
&=&
-{\cal G}_{3,qg}^{0}(s_{12})
-S_{q\to g}\Gamma^{(1)}_{gq}(x_1).
\eeqa
Here calligraphic symbols denote integrated antenna functions, e.g.,
${\cal A}_{3,qg}^{0}$ is the integrated counterpart of the
three-parton antenna $A_{3}^{0}(q,\hat{g},\hat{\bar{q}})$, while $\Ga_{ij}^{(1)}(x)$ denote the NLO mass-factorisation splitting kernels~\cite{Daleo:2006xa}.
Identity-changing antennae are accompanied by the factors $S_{g\to q}$ or $S_{q\to g}$, which correct for the fact that the degrees of freedom in $d$-dimensions are different for a gluon and quark. Explicitly they are
\beqa
S_{g\to q} &=& \frac{S_g}{S_q} = \frac{2-2\ep}{2} = 1 - \ep, \nn\\
S_{q\to g} &=& \frac{S_q}{S_g} = \frac{2}{2-2\ep} = \frac{1}{1 - \ep}. 
\eeqa
The initial-initial integrated antennae depend on the initial-state invariant $(s_{12}=2p_1\cdot p_2$), of the underlying Born kinematics.

\begin{table}[t]
\centering
\begin{tabular}{lcc}
\hline
$\sigma_{\textrm{NLO},qg}(\mu_R,\mu_F)$ & NNLOJET [pb] & {\tt Top++} [pb]  \\
\hline
$(m_t,m_t)$    & $\phantom{-}3.260(3)$   & $\phantom{-}3.2624$ \\
$(m_t,m_t/2)$  & $\phantom{-}27.980(4)$ & $\phantom{-}27.9842$\\
$(m_t,2m_t)$   & $-18.701(3)$           & $-18.7027$ \\
\hline
\end{tabular}
\caption{NLO-$qg$ channel contribution to the top-pair production cross section, in picobarns [pb], for the $qg$ channel for the renormalization and factorization scale choices 
$(\mu_R,\mu_F)=(m_t,m_t),(m_t,m_2/2),(m_t,2m_t)$.
The \textsc{NNLOJET} results are compared with the corresponding predictions prediction from {\tt Top++} at $\sqrt{s}=13$ TeV. 
The quoted NNLOJET uncertainty denoted in parenthesis on the last digit corresponds to the Monte Carlo integration error.}
\label{tab:nloqg}
\end{table}

The NLO $qg$-channel contribution described above has been implemented within the \textsc{NNLOJET} parton-level generator. This includes the real-radiation matrix elements and the corresponding subtraction terms in Eqs.~\eqref{eq:snloqg} and~\eqref{eq:tnloqg}.
To validate the implementation, we consider proton-proton collisions at a centre-of-mass energy of $\sqrt{s}=13$ TeV and employ the 
NNPDF4.0 NLO parton distribution functions~\cite{NNPDF:2021njg}. The strong coupling is evolved at two-loop order with $\alpha_s(m_Z)=0.118$, and the pole mass of the top quark is fixed to $m_t=173.3$ GeV. The numerical results for the NLO $qg$-channel contribution are presented in Table~\ref{tab:nloqg} and compared with the corresponding predictions obtained with {\tt Top++}~\cite{Czakon:2011xx}. Excellent agreement is observed between the \textsc{NNLOJET} implementation and the independent predictions obtained with {\tt Top++}.

\section {Infrared structure of top-pair production at ${\cal O}(\alpha_s^4$) in the $qg$ and $qq,qq'$ channels} 
\label{sec:ttbnnlo}
In this section we discuss the infrared structure of top-pair production at NNLO in the $qg$ and $qq,qq'$-channels within the antenna subtraction framework. For each partonic channel, the NNLO contribution is decomposed into double-real, real-virtual and double-virtual corrections according to the general structure of the partonic cross section given in Eq.~\eqref{eq.subnnlo}. In the following, we present the corresponding NNLO subtraction terms for the $qg$, $qq$ and $qq^\prime$ channels. The calculation is performed retaining the full colour dependence. For compactness, we restrict the discussion to the leading-colour subtraction terms. The subleading-colour contributions are derived from the same analytic building blocks and therefore require no additional analytic structures.
 
\subsection{NNLO corrections in the $qq$,$qq'$-channels}
By the $qq$ and $qq'$ channels we shall collectively refer to the partonic initial states $qq$, $qq'$, $\bar{q}\bar{q}$ and $\bar{q}\bar{q}'.$ The contributions with two 
antiquarks in the initial state are obtained from the corresponding quark-initiated ones by charge conjugation.

As summarised in Table~\ref{tab:ttbar_channels}, these initial states enter the perturbative expansion of top-pair production only at ${\cal O}(\alpha_s^4)$.
Their NNLO contribution consists exclusively of double-real tree-level matrix elements, with no corresponding real-virtual or double-virtual matrix-elements. Nevertheless, the integration of the double-real subtraction terms over the unresolved phase space generates initial-state collinear singularities, which are cancelled by the corresponding mass-factorisation counterterms. The NNLO calculation of these channels therefore requires the derivation of the double-real subtraction term together with its integrated counterparts entering the ${\rm d}\hat{\sigma}^{T}_{\NNLO}$ and ${\rm d}\hat{\sigma}^{U}_{\NNLO}$ subtraction terms. 
The derivation of these contributions is presented below.

\subsubsection{Double-real contribution in the $qq$ and $qq'$ channels}
The contribution of purely fermionic processes to the double-real radiation cross section for massive quark-pair production in hadronic collisions was discussed within the antenna subtraction formalism in Ref.~\cite{Abelof:2011ap}, where the colour decompositions and corresponding double-real subtraction terms were derived. In the present work we build upon these results to complete the NNLO treatment of the $qq$ and $qq'$ channels by deriving the integrated subtraction terms entering the real-virtual and double-virtual contributions and describing their implementation within \textsc{NNLOJET}. 

Owing to the presence of identical and non-identical light quarks in the initial and final states, distinct partonic subprocesses 
contribute to the hadronic cross section. The hadronic double-real contribution involving light quarks in the initial state
is given by,
\beqa
&&{\rm d}\sigma^{RR}=\int \frac{{\rm d}\xi_1}{\xi_1}
\frac{{\rm d}\xi_2}{\xi_2}\Big\{
\sum_{q}f_q(\xi_1)f_q(\xi_2)
{\rm d}\hat{\sigma}_{qq\to Q\bar{Q}qq}\nonumber\\
&&\hspace{3.1cm}+
\sum_{q\neq q'}f_q(\xi_1)f_{q'}(\xi_2){\rm d}\hat{\sigma}_{qq'\to Q\bar{Q}qq'}
\Big\}
\eeqa
with the partonic cross sections given by,
\beqa
{\rm d}\hat{\sigma}_{qq\to Q\bar{Q}qq}&=&\norm_{\NNLO}^{\:qq}
{\rm d}\Phi_4(p_Q,p_{\bar{Q}},p_q,p_q;p_1,p_2)
|{\cal M}^0_{qq\to Q\bar{Q}qq}|^2
J^{(4)}_2(p_Q,p_{\bar{Q}},p_q,p_q)
\label{eq:rrqq}
\\
{\rm d}\hat{\sigma}_{qq'\to Q\bar{Q}qq'}&=&\norm_{\NNLO}^{\:qq'}
{\rm d}\Phi_4(p_Q,p_{\bar{Q}},p_q,p_{q^{'}};p_1,p_2)
|{\cal M}^0_{qq'\to Q\bar{Q}qq'}|^2 
J^{(4)}_2(p_Q,p_{\bar{Q}},p_q,p_{q^{'}}).
\label{eq:rrqqp}
\eeqa
In Eqs.~\eqref{eq:rrqq},~\eqref{eq:rrqqp},
$J^{(4)}_2(p_Q,p_{\bar{Q}},p_q,p_{q^{'}})$ is a measurement function that ensures that the heavy quark and antiquark of momenta $p_Q$ and
$p_{\bar{Q}}$ are observed, while the two additional massless partons of momenta $p_q,p_{q^{'}}$ are treated inclusively and may either be resolved as additional jets or remain unresolved.

In order to obtain the subtraction terms, the colour decomposition of the real matrix elements present in the partonic cross sections in Eqs.~\eqref{eq:rrqq},~\eqref{eq:rrqqp} has to be performed. Since there are no gluons present in the external state, the colour structure of the
amplitude is particularly simple. All colour factors are products of Kronecker deltas in the fundamental representation. We can write the amplitude for the process 
$qq'\to Q\bar{Q}qq'$ as,
\begin{eqnarray}
&&M_6^0(5_q,\hat{1}_{\bar q},3_Q,4_{\bar Q},6_{q'},\hat{2}_{\bar q'})=
g_s^4\Bigg[
\delta_{i_3,i_1}\,\delta_{i_5,i_2}\,\delta_{i_6,i_4}\,
{\cal M}_6^{0,a}(3_Q,\hat{1}_{\bar q};;5_q,\hat{2}_{\bar q'};;6_{q'},4_{\bar Q})
\nonumber\\
&&\hspace{1.4cm}
+\delta_{i_3,i_2}\,\delta_{i_5,i_4}\,\delta_{i_6,i_1}\,
{\cal M}_6^{0,b}(3_Q,\hat{2}_{\bar q'};;5_q,4_{\bar Q};;6_{q'},\hat{1}_{\bar q})
\nonumber\\
&&\hspace{1.4cm}
-\frac{1}{N_c}\,
\delta_{i_3,i_1}\,\delta_{i_5,i_4}\,\delta_{i_6,i_2}\,
{\cal M}_6^{0,c}(3_Q,\hat{1}_{\bar q};;5_q,4_{\bar Q};;6_{q'},\hat{2}_{\bar q'})
\nonumber\\
&&\hspace{1.4cm}
-\frac{1}{N_c}\,
\delta_{i_3,i_2}\,\delta_{i_5,i_1}\,\delta_{i_6,i_4}\,
{\cal M}_6^{0,d}(3_Q,\hat{2}_{\bar q'};;5_q,\hat{1}_{\bar q};;6_{q'},4_{\bar Q})
\nonumber\\
&&\hspace{1.4cm}
-\frac{1}{N_c}\,
\delta_{i_3,i_4}\,\delta_{i_5,i_2}\,\delta_{i_6,i_1}\,
{\cal M}_6^{0,e}(3_Q,4_{\bar Q};;5_q,\hat{2}_{\bar q'};;6_{q'},\hat{1}_{\bar q})
\nonumber\\
&&\hspace{1.4cm}
+\frac{1}{N_c^2}\,
\delta_{i_3,i_4}\,\delta_{i_5,i_1}\,\delta_{i_6,i_2}\,
{\cal M}_6^{0,f}(3_Q,4_{\bar Q};;5_q,\hat{1}_{\bar q};;6_{q'},\hat{2}_{\bar q'})
\Bigg].
\label{eq:qqpcolour}
\end{eqnarray}
For the process involving two identical flavour quark pairs, $qq\to Q\bar{Q}qq$, the
colour decomposition can be obtained from Eq.~\eqref{eq:qqpcolour}
by using,
\beq
M_6^0(5_q,\hat{1}_{\bar q},3_Q,4_{\bar Q},6_{q},\hat{2}_{\bar q})=
M_6^0(5_q,\hat{1}_{\bar q},3_Q,4_{\bar Q},6_{q'},\hat{2}_{\bar q'})
-M_6^0(5_q,\hat{2}_{\bar q},3_Q,4_{\bar Q},6_{q'},\hat{1}_{\bar q'}),
\label{eq:qqid}
\eeq
where second term is obtained simply by exchanging the two 
incoming quarks.

Following  the \textsc{NNLOJET} notation introduced previously,
the matrix elements for processes involving three distinct quark-antiquark pairs are denoted by $E$-type amplitudes.
Squaring Eq.~\eqref{eq:qqpcolour} and combining it with the $2\to 4$ phase space, the appropriate overall factors and the measurement function, we can write the real radiation $qq'$-channel contribution as,
\beqa
&&{\rm d}\hat{\sigma}_{qq',NNLO}^{RR}=\nonumber\\
&&\hspace{0.6cm}\norm_{\NNLO}^{\:qq'}\int{\rm d}\Phi_4(p_3,p_4,p_5,p_6;p_1,p_2)
\Big\{N_c\;
E_{0}^{0}(5_q,\hat{1}_{\bar q},3_Q,4_{\bar Q},6_{q'},\hat{2}_{\bar q'})\nonumber\\
&&\hspace{2.9cm}+\frac{1}{N_c}\tilde{E}_{0}^{0}(5_q,\hat{1}_{\bar q},3_Q,4_{\bar Q},6_{q'},\hat{2}_{\bar q'})\Big\}J^{(4)}_2(p_3,p_4,p_5,p_6),
\label{eq:nnloRRqqp2}
\eeqa
where,
\beqa
E_{0}^{0}(5_q,\hat{1}_{\bar q},3_Q,4_{\bar Q},6_{q'},\hat{2}_{\bar q'})&=&
E_{0}^{0,a}(3_Q,\hat{1}_{\bar q};;5_q,\hat{2}_{\bar q'};;6_{q'},4_{\bar Q})+
E_{0}^{0,b}(3_Q,\hat{2}_{\bar q'};;5_q,4_{\bar Q};;6_{q'},\hat{1}_{\bar q})\nonumber\\
\tilde{E}_{0}^{0}(5_q,\hat{1}_{\bar q},3_Q,4_{\bar Q},6_{q'},\hat{2}_{\bar q'})&=&
E_{0}^{0,c}(3_Q,\hat{1}_{\bar q};;5_q,4_{\bar Q};;6_{q'},\hat{2}_{\bar q'})
+
E_{0}^{0,d}(3_Q,\hat{2}_{\bar q'};;5_q,\hat{1}_{\bar q};;6_{q'},4_{\bar Q})\nonumber\\
&+&
E_{0}^{0,e}(3_Q,4_{\bar Q};;5_q,\hat{2}_{\bar q'};;6_{q'},\hat{1}_{\bar q})
-3\,
E_{0}^{0,f}(3_Q,4_{\bar Q};;5_q,\hat{1}_{\bar q};;6_{q'},\hat{2}_{\bar q'}).\nonumber
\eeqa
The quantities $E_{0}^{0,a}$, \ldots, $E_{0}^{0,f}$ denote the squared partial amplitudes
corresponding to the colour decomposition, e.g.,
\beq
E_{0}^{0,a}(3_Q,\hat{1}_{\bar q};;5_q,\hat{2}_{\bar q'};;6_{q'},4_{\bar Q})
=
\left|
{\cal M}_6^{0,a}(3_Q,\hat{1}_{\bar q};;5_q,\hat{2}_{\bar q'};;6_{q'},4_{\bar Q})
\right|^2,
\eeq
with analoguous definitions for the remaining colour-stripped partial amplitudes. The overall normalisation factor is given by
\beq
\norm_{\NNLO}^{\,qq'}=
{\cal N}_{\rm LO}^{q\bar{q}}
\left(\frac{\alpha_s}{2\pi}\right)^2
\frac{\bar C(\epsilon)^2}{C(\epsilon)^2}.
\eeq

Finally, using~\eqref{eq:qqid}, the double-real partonic cross section for the identical-flavour process $qq\to QQ qq$ is given by,
\beqa
&&{\rm d}\hat{\sigma}_{qq,NNLO}^{RR}=
\norm_{\NNLO}^{\:qq}\int{\rm d}\Phi_4(p_3,p_4,p_5,p_6;p_1,p_2)
\Big\{\nonumber\\
&&\hspace{1.6cm} N_c\;\Big(
E_{0}^{0}(5_q,\hat{1}_{\bar q},3_Q,4_{\bar Q},6_{q'},\hat{2}_{\bar q'})
+
E_{0}^{0}(5_q,\hat{2}_{\bar q},3_Q,4_{\bar Q},6_{q'},\hat{1}_{\bar q'})
\Big)
\nonumber\\
&&\hspace{1.4cm}+\;F_0^{0}(5_q,\hat{1}_{\bar q},3_Q,4_{\bar Q},6_{q},\hat{2}_{\bar q})\nonumber\\
&&\hspace{1.4cm}+\frac{1}{N_c}\Big(
\tilde{E}_{0}^{0}(5_q,\hat{1}_{\bar q},3_Q,4_{\bar Q},6_{q'},\hat{2}_{\bar q'})
+
\tilde{E}_{0}^{0}(5_q,\hat{2}_{\bar q},3_Q,4_{\bar Q},6_{q'},\hat{1}_{\bar q'})
\Big)\nonumber\\
&&\hspace{1.4cm}+\frac{1}{N_c^2}\;\tilde{F}_0^{0}(5_q,\hat{1}_{\bar q},3_Q,4_{\bar Q},6_{q},\hat{2}_{\bar q})
\Big\}J^{(4)}_2(p_3,p_4,p_5,p_6).
\label{eq:nnloRRqq2}
\eeqa
With respect to the non-identical flavour process we observe the appearance of two additional colour structures that contain interferences of partial amplitudes,
\beqa
F_0^{0}(5_q,\hat{1}_{\bar q},3_Q,4_{\bar Q},6_{q},\hat{2}_{\bar q})
&=& 2\re(\cm_6(3_Q,\hat{2}_{\bar q'};;5_q,4_{\bar Q};;6_{q'},\hat{1}_{\bar q})\cm_6(3_Q,\hat{2}_{\bar q'};;5_q,4_{\bar Q};;6_{q'},\hat{1}_{\bar q})^{\dagger})\nonumber\\
&+& 2\re(\cm_6(3_Q,\hat{2}_{\bar q'};;5_q,\hat{1}_{\bar q};;6_{q'},4_{\bar Q})\cm_6(3_Q,\hat{2}_{\bar q'};;5_q,\hat{1}_{\bar q};;6_{q'},4_{\bar Q})^{\dagger})\nonumber\\
&+& 2\re(\cm_6(3_Q,\hat{1}_{\bar q};;5_q,4_{\bar Q};;6_{q'},\hat{2}_{\bar q'})\cm_6(3_Q,\hat{1}_{\bar q};;5_q,4_{\bar Q};;6_{q'},\hat{2}_{\bar q'})^{\dagger})\nonumber\\
&+&2\re(\cm_6(3_Q,\hat{1}_{\bar q};;5_q,\hat{2}_{\bar q'};;6_{q'},4_{\bar Q})\cm_6(3_Q,\hat{1}_{\bar q};;5_q,\hat{2}_{\bar q'};;6_{q'},4_{\bar Q})^{\dagger})\nonumber\\
&-& 2\re(\cm_6(3_Q,4_{\bar Q};;5_q,\hat{1}_{\bar q};;6_{q'},\hat{2}_{\bar q'})\cm_6(3_Q,4_{\bar Q};;5_q,\hat{2}_{\bar q};;6_{q'},\hat{1}_{\bar q})^{\dagger})\nonumber\\
\label{eq:F00X}
&&\\
\tilde{F}_0^{0}(5_q,\hat{1}_{\bar q},3_Q,4_{\bar Q},6_{q},\hat{2}_{\bar q})
&=&6\re(\cm_6(3_Q,4_{\bar Q};;5_q,\hat{1}_{\bar q};;6_{q'},\hat{2}_{\bar q'})
\cm_6(3_Q,4_{\bar Q};;5_q,\hat{1}_{\bar q};;6_{q'},\hat{2}_{\bar q'})^{\dagger})
\nonumber\\
&-&
2\re(\cm_6(3_Q,4_{\bar Q};;5_q,\hat{1}_{\bar q};;6_{q'},\hat{2}_{\bar q'})
\cm_6(3_Q,4_{\bar Q};;5_q,\hat{1}_{\bar q};;6_{q'},\hat{2}_{\bar q'})^{\dagger})
\nonumber\\
&-&
2\re(\cm_6(3_Q,4_{\bar Q};;5_q,\hat{2}_{\bar q'};;6_{q'},\hat{1}_{\bar q})
\cm_6(3_Q,4_{\bar Q};;5_q,\hat{2}_{\bar q'};;6_{q'},\hat{1}_{\bar q})^{\dagger}).\nonumber\\
\eeqa
The overall normalisation factor is given by
\beq
\norm_{\NNLO}^{\,qq}=\norm_{\NNLO}^{\,qq'}\,\frac{1}{2!},
\eeq
where the additional symmetry factor of $1/2!$ accounts for identical particles in the final state. 

Having established the colour decomposition of the matrix elements contributing to the $qq'$ and $qq$-channels, we now turn to the derivation of the corresponding antenna subtraction terms. As discussed previously, quasi-collinear limits involving the massive quarks are not subtracted explicitly. The full matrix element contribution therefore exhibits only collinear unresolved configurations involving the massless quarks. The double-unresolved singularities arise in the initial-state triple-collinear limits $\hat{1}_{\bar q}||5_q||6_{q'}$
and $\hat{2}_{\bar q'}||5_q||6_{q'}$, together with the double-collinear configuration $\hat{1}_{\bar q}||5_q+\hat{2}_{\bar q'}||6_{q'}$. In addition, single-collinear limits occur when either
$\hat{1}_{\bar q}||5_q$ or $\hat{2}_{\bar q'}||6_{q'}$.
No additional soft or double-soft singularities occur beyond the collinear configurations listed above. The subtraction term is derived by associating each of these unresolved limits with the appropriate antenna function.

The triple-collinear configurations are captured by the massless initial-initial four-parton antenna $B_4^0$~\cite{Gehrmann-DeRidder:2012too}, while the remaining single-and double collinear limits are described by three-parton $E_3^0$~\cite{Abelof:2011jv} antennae involving one massive and one massless radiator.
The corresponding subtraction term is obtained by combining these antenna functions with reduced matrix-elements so as to reproduce the real-real matrix element squared in all unresolved limits. At leading colour it is given as , ${\rm d}\hat{\sigma}_{qq',\NNLO}^{S,l.c.}$ by
\beqa
&&\hspace{-3.0cm}{\rm d}\hat{\sigma}_{qq',\NNLO}^{S,l.c.}=\norm_{\NNLO}^{\,qq'}\,N_c\,
{\rm d}\Phi_4(p_3,p_4,p_5,p_6;p_1,p_2)\Big\{\nonumber\\
\quad&-&\frac{1}{2}E_{3}^{0}(3_Q,5_q,\hat{1}_q)\;C_1^{0,X}(6_{q'},\hat{2}_{q'},(35)_Q,4_{\bar{Q}},\hat{\bar{1}}_g)\nonumber\\
\quad&-&\frac{1}{2}E_{3}^{0}(4_{\bar{Q}},5_q,\hat{1}_q)\;C_1^{0,X}(6_{q'},\hat{2}_{q'},3_Q,(45)_Q,\hat{\bar{1}}_g)\nonumber\\
\quad&-&\frac{1}{2}E_{3}^{0}(3_Q,6_{q'},\hat{2}_{q'})\;C_1^{0,X}(5_q,\hat{1}_q,(36)_Q,4_{\bar{Q}},\hat{\bar{2}}_g)\nonumber\\
\quad&-&\frac{1}{2}E_{3}^{0}(4_{\bar{Q}},6_{q'},\hat{2}_{q'})\;C_1^{0,X}(5_q,\hat{1}_q,3_Q,(46)_Q,\hat{\bar{2}}_g)\nonumber\\
&&\nonumber\\
\quad&+&B_4^0(5_q,\hat{2}_{q'},6_{q'},\hat{1}_q)C_0^0(\hat{\bar{1}}_q,\tilde{3}_Q,\tilde{4}_Q,\hat{\bar{2}}_q)\nonumber\\
\quad&-&\frac{1}{2}E_{3}^{0}(3_Q,6_{q'},\hat{2}_{q'})A_3^0(5_q,\hat{\bar{2}}_g,\hat{1}_q)C_0^0(\hat{\bar{1}}_q,\widetilde{(36)}_Q,\tilde{4}_Q,\hat{\bar{\bar{2}}}_q)\nonumber\\
\quad&-&\frac{1}{2}E_{3}^{0}(4_{\bar{Q}},6_{q'},\hat{2}_{q'})A_3^0(5_q,\hat{\bar{2}}_g,\hat{1}_q)C_0^0(\hat{\bar{1}}_q,\tilde{3}_Q,\widetilde{(46)}_Q,\hat{\bar{\bar{2}}}_q)\nonumber\\
&&\nonumber\\
\quad&+&B_4^0(6_{q'},\hat{1}_q,5_q,\hat{2}_{q'})C_0^0(\hat{\bar{1}}_q,\tilde{3}_Q,\tilde{4}_Q,\hat{\bar{2}}_q)\nonumber\\
\quad&-&\frac{1}{2}E_{3}^{0}(3_Q,5_q,\hat{1}_q)A_3^0(6_{q'},\hat{\bar{1}}_g,\hat{2}_{q'})C_0^0(\hat{\bar{\bar{1}}}_q,\widetilde{(35)}_Q,\tilde{4}_Q,\hat{\bar{2}}_q)\nonumber\\
\quad&-&\frac{1}{2}E_{3}^{0}(4_{\bar{Q}},5_q,\hat{1}_q)A_3^0(6_{q'},\hat{\bar{1}}_g,\hat{2}_{q'})C_0^0(\hat{\bar{\bar{1}}}_q,\tilde{3}_Q,\widetilde{(45)}_Q,\hat{\bar{2}}_q)\nonumber\\
&&\nonumber\\
\quad&-&\frac{1}{2}E_{3}^{0}(3_Q,5_q,\hat{1}_q)E_3^0(4_{\bar{Q}},6_{q'},\hat{2}_{q'})\;B_2^0(\widetilde{(35)}_Q,\hat{\bar{1}}_g,\hat{\bar{2}}_g,\widetilde{(46)}_Q)\nonumber\\
\quad&-&\frac{1}{2}E_{3}^{0}(3_Q,5_q,\hat{1}_q)E_3^0(4_{\bar{Q}},6_{q'},\hat{2}_{q'})\;B_2^0(\widetilde{(35)}_Q,\hat{\bar{2}}_g,\hat{\bar{1}}_g,\widetilde{(46)}_Q)\nonumber\\
\quad&-&\frac{1}{2}E_{3}^{0}(3_Q,6_{q'},\hat{2}_{q'})E_3^0(4_{\bar{Q}},5_q,\hat{1}_q)\;B_2^0(\widetilde{(36)}_Q,\hat{\bar{1}}_g,\hat{\bar{2}}_g,\widetilde{(45)}_Q)\nonumber\\
\quad&-&\frac{1}{2}E_{3}^{0}(3_Q,6_{q'},\hat{2}_{q'})E_3^0(4_{\bar{Q}},5_q,\hat{1}_q)\;B_2^0(\widetilde{(36)}_Q,\hat{\bar{2}}_g,\hat{\bar{1}}_g,\widetilde{(45)}_Q)\Big\}
\label{eq:qqpS}
\eeqa
where the colour stripped C-type reduced matrix elements 
have been grouped into the combination,
\beq
C_1^{0,X}(5_q,\hat{1}_q,3_Q,4_{\bar{Q}},\hat{2}_g)=
C_1^{0,a}(3_Q,\hat{2}_g,\hat{1}_q;;5_q,4_{\bar{Q}})+
C_1^{0,b}(3_Q,\hat{1}_q;;5_q,\hat{2}_g,4_{\bar{Q}})\,.
\eeq
With the double real subtraction term for the non-identical flavour case established, the leading-colour contributions to the identical-flavour process follows straightforwardly. The only genuinely new colour structures in the identical-flavour channel are therefore the interference contributions $F_0^{0}$ and $\tilde{F}_0^{0}$.

In the following we discuss explicitly only the subtraction of the colour
structure $F_0^{0}$, defined in 
Eq.~\eqref{eq:F00X}. In contrast to the squared partial amplitudes, it exhibits a much simpler infrared structure. Its singular behaviour is restricted to the initial-state triple-collinear limits $\hat{1}_{\bar q}||5_q||6_q$
and $\hat{2}_{\bar q}||5_q||6_q$, since all single-collinear and double-collinear singularities cancel when the partial amplitudes are interfered. The triple collinear limits limit are captured entirely by an initial-initial four-parton antenna $C_4^0$. 
The corresponding subtraction term is given by
\beqa
&&\hspace{-3.0cm}{\rm d}\hat{\sigma}_{qq,\NNLO}^{S,l.c.}=\norm_{\NNLO}^{\,qq}
{\rm d}\Phi_4(p_3,p_4,p_5,p_6;p_1,p_2)\Big\{\nonumber\\
\quad&-&2\,C_4^0(\hat{1}_{\bar{q}},\hat{2}_{\bar{q}},5_q,6_q)
C_0^0(\hat{\bar{1}}_q,\tilde{3}_Q,\tilde{4}_Q,\hat{\bar{2}}_q)\nonumber\\
\quad&-&2\,C_4^0(\hat{2}_{\bar{q}},\hat{1}_{\bar{q}},5_q,6_q)
C_0^0(\hat{\bar{1}}_q,\tilde{3}_Q,\tilde{4}_Q,\hat{\bar{2}}_q)\Big\}.
\eeqa

Having completed the discussion of the leading-colour subtraction terms, we note that the corresponding subleading-colour contributions have also been derived and implemented in the \textsc{NNLOJET} framework. They are fully included in the numerical results presented in Section~\ref{sec:nnlonumerics}, but are not displayed explicitly here in order to keep the presentation compact.

\subsubsection{Real-virtual contribution in the $qq$ and $qq'$ channels}
Although the $qq'$ and $qq$-channels do not receive genuine real-virtual matrix-element contributions, the integrated double-real subtraction terms above generate contributions to ${\rm d}\hat{\sigma}^T_{\NNLO}$. After cancellation of the explicit initial-state collinear poles against the mass-factorisation counterterms, the resulting reduced matrix elements still describe $Q\bar{Q}$+jet production and therefore retain single-unresolved singularities that must be subtracted locally.

Since these reduced matrix elements correspond to squared tree-level amplitudes, their unresolved limits are subtracted using tree level three-parton massless antennae. Initial-state quark-gluon collinear configurations are captured by initial-initial $A_3^0$
antennae~\cite{Daleo:2006xa}, while the remaining quark-quark collinear limits are captured by massive $E_3^0$~\cite{Abelof:2011jv} antennae involving one massive and one massless radiator. The resulting leading-colour contribution to the
${\rm d}\hat{\sigma}^T_{\NNLO}$ subtraction term is then given by,
\beqa
&&\hspace{-2.0cm}{\rm d}\hat{\sigma}_{qq',\NNLO}^{T,l.c.}=\norm_{\NNLO}^{\,qq'}
C(\epsilon)\,N_c\,
{\rm d}\Phi_3(p_3,p_4,p_5;p_1,p_2)\Big\{\nonumber\\
\quad&-&\frac{1}{2}({\cal E}_{3,q}^{0,m}(s_{13})-\Gamma_{gq}(x_1))C_1^{0,X}(5_q,\hat{2}_q,3_Q,4_{\bar{Q}},\hat{1}_g)\nonumber\\
\quad&-&\frac{1}{2}({\cal E}_{3,q}^{0,m}(s_{14})-\Gamma_{gq}(x_1))C_1^{0,X}(5_q,\hat{2}_q,3_Q,4_{\bar{Q}},\hat{1}_g)\nonumber\\
\quad&-&\frac{1}{2}({\cal E}_{3,q}^{0,m}(s_{23})-\Gamma_{gq}(x_2))C_1^{0,X}(5_q,\hat{1}_q,3_Q,4_{\bar{Q}},\hat{2}_g)\nonumber\\
\quad&-&\frac{1}{2}({\cal E}_{3,q}^{0,m}(s_{24})-\Gamma_{gq}(x_2))C_1^{0,X}(5_q,\hat{1}_q,3_Q,4_{\bar{Q}},\hat{2}_g)\nonumber\\
&&\nonumber\\
\quad&-&\frac{1}{2}A_3^0(5_q,\hat{2}_g,\hat{1}_q)({\cal E}_{3,q}^{0,m}(s_{23})+{\cal E}_{3,q}^{0,m}(s_{24})-\Gamma_{gq}(x_2))C_0^0(\hat{\bar{1}}_q,\tilde{3}_Q,\tilde{4}_{\bar{Q}},\hat{\bar{2}}_q)\nonumber\\
\quad&-&\frac{1}{2}A_3^0(5_q,\hat{1}_g,\hat{2}_q)({\cal E}_{3,q}^{0,m}(s_{13})+{\cal E}_{3,q}^{0,m}(s_{14})-\Gamma_{gq}(x_1))C_0^0(\hat{\bar{1}}_q,\tilde{3}_Q,\tilde{4}_{\bar{Q}},\hat{\bar{2}}_q)\nonumber\\
&&\nonumber\\
\quad&-&\frac{1}{2}E_{3}^{0}(3_Q,5_{q'},\hat{2}_{q'})({\cal E}_{3,q}^{0,m}(s_{14})-\Gamma_{gq}(x_1))B_2^0(\widetilde{(35)}_Q,\hat{1}_g,\hat{\bar{2}}_g,4_{\bar{Q}})\nonumber\\
\quad&-&\frac{1}{2}E_{3}^{0}(3_Q,5_{q'},\hat{2}_{q'})({\cal E}_{3,q}^{0,m}(s_{14})-\Gamma_{gq}(x_1))B_2^0(\widetilde{(35)}_Q,\hat{\bar{2}}_g,\hat{1}_g,4_{\bar{Q}})\nonumber\\
\quad&-&\frac{1}{2}E_{3}^{0}(4_Q,5_{q'},\hat{2}_{q'})({\cal E}_{3,q}^{0,m}(s_{13})-\Gamma_{gq}(x_1))B_2^0(3_Q,\hat{1}_g,\hat{\bar{2}}_g,\widetilde{(45)}_{\bar{Q}})\nonumber\\
\quad&-&\frac{1}{2}E_{3}^{0}(4_Q,5_{q'},\hat{2}_{q'})({\cal E}_{3,q}^{0,m}(s_{13})-\Gamma_{gq}(x_1))B_2^0(3_Q,\hat{\bar{2}}_g,\hat{1}_g,\widetilde{(45)}_{\bar{Q}})\nonumber\\
\quad&-&\frac{1}{2}E_{3}^{0}(3_Q,5_{q'},\hat{1}_{q'})({\cal E}_{3,q}^{0,m}(s_{24})-\Gamma_{gq}(x_2))B_2^0(\widetilde{(35)}_Q,\hat{\bar{1}}_g,\hat{2}_g,4_{\bar{Q}})\nonumber\\
\quad&-&\frac{1}{2}E_{3}^{0}(3_Q,5_{q'},\hat{1}_{q'})({\cal E}_{3,q}^{0,m}(s_{24})-\Gamma_{gq}(x_2))B_2^0(\widetilde{(35)}_Q,\hat{2}_g,\hat{\bar{1}}_g,4_{\bar{Q}})\nonumber\\
\quad&-&\frac{1}{2}E_{3}^{0}(4_Q,5_{q'},\hat{1}_{q'})({\cal E}_{3,q}^{0,m}(s_{23})-\Gamma_{gq}(x_2))B_2^0(3_Q,\hat{\bar{1}}_g,\hat{2}_g,\widetilde{(45)}_{\bar{Q}})\nonumber\\
\quad&-&\frac{1}{2}E_{3}^{0}(4_Q,5_{q'},\hat{1}_{q'})({\cal E}_{3,q}^{0,m}(s_{23})-\Gamma_{gq}(x_2))B_2^0(3_Q,\hat{2}_g,\hat{\bar{1}}_g,\widetilde{(45)}_{\bar{Q}})\,\Big\},
\eeqa
where calligraphic symbols denote integrated antenna functions, e.g.,
${\cal E}_{3,q}^{0,m}$ is the integrated form of the corresponding massive $E_3^0(Q,q',\hat{q}')$ antenna~\cite{Abelof:2011jv}.

\subsubsection{Double-virtual contribution in the $qq$ and $qq'$ channels}
The final contribution to the NNLO subtraction for the $qq'$ and $qq$-channels is contained in 
${\rm d}\hat{\sigma}^U_{\NNLO}$. Although no genuine double-virtual matrix elements contribute to these partonic channels, the integration of the double-real and real-virtual subtraction terms generates a non-vanishing $U$-type contribution. This contribution contains explicit infrared singularities, namely initial-state collinear singularities, which are cancelled analytically by the corresponding second-order mass-factorisation counterterms. The resulting subtraction term depends only on Born-level matrix elements multiplied by integrated antenna functions and mass-factorisation kernels. The leading-colour contribution to the
$qq'$-channel double-virtual subtraction reads,
\beqa
&&\hspace{-2.0cm}{\rm d}\hat{\sigma}_{qq',\NNLO}^{U,l.c.}=\norm_{\NNLO}^{\,qq'}
C(\epsilon)^2\,N_c\,
{\rm d}\Phi_2(p_3,p_4;p_1,p_2)\Big\{\nonumber\\
\quad&-&\Big({\cal B}_{4,qq'}^{0}(s_{12})+S_{q\to g}\Gamma^1_{gq}(x_2)\otimes{\cal A}_{3,qg}^0(s_{12})+\frac{1}{2}S_{q\to g}\Gamma^1_{gq}(x_2)\otimes S_{g\to q}\Gamma^1_{qg}(x_2)\nonumber\\
&\quad&-\Gamma^2_{qq'}(x_2)\Big)C_0^{0}(\hat{1}_q,3_Q,4_{\bar{Q}},\hat{2}_{\bar{q}})\nonumber\\
\quad&-&\Big({\cal B}_{4,q'q}^{0}(s_{12})+S_{q\to g}\Gamma^1_{gq}(x_1)\otimes{\cal A}_{3,gq}^0(s_{12})+\frac{1}{2}S_{q\to g}\Gamma^1_{gq}(x_1)\otimes S_{g\to q}\Gamma^1_{qg}(x_1)\nonumber\\
&\quad&-\Gamma^2_{qq'}(x_1)\Big)C_0^{0}(\hat{1}_q,3_Q,4_{\bar{Q}},\hat{2}_{\bar{q}})\nonumber\\
\quad&-&\frac{1}{2}\Big({\cal E}_{3,q'}^{0,m}(s_{23})\otimes{\cal E}_{3,q'}^{0,m}(s_{14})+{\cal E}_{3,q'}^{0,m}(s_{24})\otimes{\cal E}_{3,q'}^{0,m}(s_{13})\nonumber\\
\quad&&-{\cal E}_{3,q'}^{0,m}(s_{23})\otimes S_{q\to g}\Gamma^1_{gq}(x_1)-{\cal E}_{3,q'}^{0,m}(s_{24})\otimes S_{q\to g}\Gamma^1_{gq}(x_1)\nonumber\\
\quad&&-{\cal E}_{3,q'}^{0,m}(s_{13})\otimes S_{q\to g}\Gamma^1_{gq}(x_2)-{\cal E}_{3,q'}^{0,m}(s_{14})\otimes S_{q\to g}\Gamma^1_{gq}(x_2)\nonumber\\
\quad&&+2\,S_{q\to g}^2\;\Gamma^1_{gq}(x_1)\otimes \Gamma^1_{gq}(x_2)\Big)B_2^0(3_Q,\hat{1}_g,\hat{2}_g,4_{\bar{Q}})\nonumber\\
\quad&-&\frac{1}{2}\Big({\cal E}_{3,q'}^{0,m}(s_{23})\otimes{\cal E}_{3,q'}^{0,m}(s_{14})+{\cal E}_{3,q'}^{0,m}(s_{24})\otimes{\cal E}_{3,q'}^{0,m}(s_{13})\nonumber\\
\quad&&-{\cal E}_{3,q'}^{0,m}(s_{23})\otimes S_{q\to g}\Gamma^1_{gq}(x_1)-{\cal E}_{3,q'}^{0,m}(s_{24})\otimes S_{q\to g}\Gamma^1_{gq}(x_1)\nonumber\\
\quad&&-{\cal E}_{3,q'}^{0,m}(s_{13})\otimes S_{q\to g}\Gamma^1_{gq}(x_2)-{\cal E}_{3,q'}^{0,m}(s_{14})\otimes S_{q\to g}\Gamma^1_{gq}(x_2)\nonumber\\
\quad&&+2\,S_{q\to g}^2\;\Gamma^1_{gq}(x_1)\otimes \Gamma^1_{gq}(x_2)\Big)B_2^0(3_Q,\hat{2}_g,\hat{1}_g,4_{\bar{Q}})\Big\}\,.
\eeqa

The structure of the subtraction term can be made more transparent by introducing integrated identity-changing dipoles. In particular, 
the combination of integrated four-parton antennae and mass-factorisation kernels appearing in the first two terms above are precisely those entering the double integrated identity-changing dipole
 $\mathcal{J}_{2,q\to q'}^{(2)}(\bar{1}_{q\to q^{'}},\bar{2}_{\bar{q}})$
introduced in Ref.~\cite{Chen:2022clm}. The remaining terms naturally factorise into products of single integrated initial-final and final-initial identity-changing dipoles involving massive radiators,
\beqa
\mathcal{J}_{2,q\to g}^{(1)}(i_Q,\bar{1}_g)&=&
{\cal E}_{3,q'}^{0,m}(s_{\bar{1}i})-S_{q\to g}\Gamma^{(1)}_{gq}(x_1)
\\
\mathcal{J}_{2,q\to g}^{(1)}(i_Q,\bar{2}_g)&=&
{\cal E}_{3,q'}^{0,m}(s_{\bar{2}i})-S_{q\to g}\Gamma^{(1)}_{gq}(x_2).
\eeqa
Introducing these newly defined operators, the U-type subtraction term can be written in the more compact form,
\beqa
&&\hspace{-2.0cm}{\rm d}\hat{\sigma}_{qq',\NNLO}^{U,l.c.}=\norm_{\NNLO}^{\,qq'}
C(\epsilon)^2\,N_c\,
{\rm d}\Phi_2(p_3,p_4;p_1,p_2)\Big\{\nonumber\\
\quad&&-\mathcal{J}_{2,q\to q'}^{(2)}(\bar{1}_{q\to q^{'}},\bar{2}_{\bar{q}})C_0^{0}(\hat{1}_q,3_Q,4_{\bar{Q}},\hat{2}_{\bar{q}})\nonumber\\
\quad&&-\mathcal{J}_{2,q\to q'}^{(2)}(\bar{2}_{q\to q^{'}},\bar{1}_{\bar{q}})C_0^{0}(\hat{1}_q,3_Q,4_{\bar{Q}},\hat{2}_{\bar{q}})\nonumber\\
\quad&&-\frac{1}{2}\mathcal{J}_{2,q\to g}^{(1)}(3_Q,\bar{2}_g)
\otimes \mathcal{J}_{2,q\to g}^{(1)}(4_Q,\bar{1}_g)
B_2^0(3_Q,\hat{1}_g,\hat{2}_g,4_{\bar{Q}})\nonumber\\
\quad&&-\frac{1}{2}\mathcal{J}_{2,q\to g}^{(1)}(4_Q,\bar{2}_g)
\otimes \mathcal{J}_{2,q\to g}^{(1)}(3_Q,\bar{1}_g)
B_2^0(3_Q,\hat{1}_g,\hat{2}_g,4_{\bar{Q}})\nonumber\\
\quad&&-\frac{1}{2}\mathcal{J}_{2,q\to g}^{(1)}(3_Q,\bar{1}_g)
\otimes \mathcal{J}_{2,q\to g}^{(1)}(4_Q,\bar{2}_g)
B_2^0(3_Q,\hat{2}_g,\hat{1}_g,4_{\bar{Q}})\nonumber\\
\quad&&-\frac{1}{2}\mathcal{J}_{2,q\to g}^{(1)}(4_Q,\bar{1}_g)
\otimes \mathcal{J}_{2,q\to g}^{(1)}(3_Q,\bar{2}_g)
B_2^0(3_Q,\hat{2}_g,\hat{1}_g,4_{\bar{Q}})\Big\}.
\label{eq:Uqqp}
\eeqa
This compact form highlights the universal structure of the subtraction term in terms of integrated identity-changing dipoles. 
Its implementation requires the analytic evaluation of the convolution products $\mathcal{J}_2^{(1)}\otimes \mathcal{J}_2^{(1)}$ appearing
in~\eqref{eq:Uqqp}. These convolutions are derived here for the first time and constitute a new ingredient of the antenna formalism for processes involving massive fermions. Together with the corresponding second-order mass-factorisation counterterms, they ensure the analytic cancellation of all explicit infrared and initial-state collinear poles at NNLO in this channel. The corresponding numerical validation of the complete implementation of the non-identical and identical channel contributions to top pair production is presented in Section~\ref{sec:nnlonumerics}.

\subsection{NNLO corrections in the $qg$-channel}
\label{sec:nnloqgS}
As summarised in Table~\ref{tab:ttbar_channels}, these initial states first contribute to top-quark pair production at ${\cal O}(\alpha_s^3)$. At ${\cal O}(\alpha_s^4)$ they receive contributions from both double-real and real-virtual matrix elements, while no double-virtual matrix element is present. The calculation has been carried out retaining the full colour dependence and is presented here for the first time.

\subsubsection{Double-real contribution in the $qg$-channel}
By the double-real contribution to the $qg$-channel we
collectively include the partonic initial states
$qg$, $\bar{q}g$, $gq$ and $g\bar{q}$, with the $\bar{q}g$ contributions
obtained from $qg$ by charge conjugation. 
Similarly, the  $gq$ and $g\bar{q}$ contributions follow by exchanging the incoming parton momenta. The corresponding hadronic contribution in this channel can then be written as,
\begin{eqnarray}
{\rm d}\sigma_{qg}^{RR}
&=&
\int\frac{{\rm d}\xi_1}{\xi_1}
\frac{{\rm d}\xi_2}{\xi_2}
\sum_q
\Big[
\big(f_q(\xi_1)+f_{\bar q}(\xi_1)\big)f_g(\xi_2)\,
{\rm d}\hat\sigma_{qg\to Q\bar Q qg}
\nonumber\\
&&\hspace{2.5cm}
+f_g(\xi_1)\big(f_q(\xi_2)+f_{\bar q}(\xi_2)\big)\,
{\rm d}\hat\sigma_{gq\to Q\bar Q qg}
\Big].
\end{eqnarray}

In the following we discuss explicitly only the $qg$ initial state.
The partonic cross section is given by
\beq
{\rm d}\hat{\sigma}_{qg\to Q\bar{Q}qg}=\norm_{\NNLO}^{\:qg}
{\rm d}\Phi_4(p_Q,p_{\bar{Q}},p_q,p_g;p_1,p_2)\,
|{\cal M}^0_{qg\to Q\bar{Q}qg}|^2\,
J^{(4)}_2(p_Q,p_{\bar{Q}},p_q,p_g).
\label{eq:rrqg}
\eeq
In turn, the tree-level amplitude for the process $qg\to Q\bar{Q}qg$ admits the following colour decomposition,
\beqa
&&\hspace{-0.2in}
M^0_{q_1 g_2 \rightarrow Q_3 \bar{Q}_4 q_5 g_6}
=
2\,g_s^4
\sum_{\{2,6\}\in \sigma_2}
\bigg[
(T^{a_{\sigma(2)}}T^{a_{\sigma(6)}})_{i_3i_1}\delta_{i_5i_4}\,
\cm_6(\Q{3},\sigma(\hat{2})_g,\sigma(6)_g,\qi{1};;\q{5},\Qb{4})
\nonumber\\
&&\hspace{1.7in}
+(T^{a_{\sigma(2)}})_{i_3i_1}(T^{a_{\sigma(6)}})_{i_5i_4}\,
\cm_6(\Q{3},\sigma(\hat{2})_g,\qi{1};;\q{5},\sigma(6)_g,\Qb{4})
\nonumber\\
&&\hspace{1.7in}
+\delta_{i_3i_1}(T^{a_{\sigma(2)}}T^{a_{\sigma(6)}})_{i_5i_4}\,
\cm_6(\Q{3},\qi{1};;\q{5},\sigma(\hat{2})_g,\sigma(6)_g,\Qb{4})
\nonumber\\
&&\hspace{1.7in}
-\frac{1}{N_c}(T^{a_{\sigma(2)}}T^{a_{\sigma(6)}})_{i_3i_4}\delta_{i_5i_1}\,
\cm_6(\Q{3},\sigma(\hat{2})_g,\sigma(6)_g,\Qb{4};;\q{5},\qi{1})
\nonumber\\
&&\hspace{1.7in}
-\frac{1}{N_c}(T^{a_{\sigma(2)}})_{i_3i_4}(T^{a_{\sigma(6)}})_{i_5i_1}\,
\cm_6(\Q{3},\sigma(\hat{2})_g,\Qb{4};;\q{5},\sigma(6)_g,\qi{1})
\nonumber\\
&&\hspace{1.7in}
-\frac{1}{N_c}\delta_{i_3i_4}(T^{a_{\sigma(2)}}T^{a_{\sigma(6)}})_{i_5i_1}\,
\cm_6(\Q{3},\Qb{4};;\q{5},\sigma(\hat{2})_g,\sigma(6)_g,\qi{1})
\bigg],\nonumber\\
\label{eq:qgcolour}
\eeqa
where $\sigma_2$ denotes the set of two permutations of the two gluons
$\{\hat{2}_g,6_g\}$. 
Squaring Eq.~\eqref{eq:qgcolour} and combining it with the $2\to 4$ phase space, the appropriate overall factors and the measurement function, we can write the real radiation $qg$-channel contribution as,
\beqa
&&{\rm d}\hat{\sigma}_{qg,NNLO}^{RR}=
\norm_{\NNLO}^{\:qg}\int{\rm d}\Phi_4(p_3,p_4,p_5,p_6;p_1,p_2)
\Big\{N_c^2\;
\Big(C_2^{0,X_1}(3_Q,\hat{2}_g,6_g,\hat{1}_q;;5_q,4_Q)\nonumber\\
&&\hspace{1.0cm}
+\:C_2^{0,X_2}(3_Q,\hat{1}_q;;5_q,\hat{2}_g,6_g,4_Q)
+\:C_2^{0,X_3}(3_Q,\hat{2}_g,\hat{1}_q;;5_q,6_g,4_Q)\nonumber\\
&&\hspace{1.0cm}
+\:C_2^{0,X_1}(3_Q,6_g,\hat{2}_g,\hat{1}_q;;5_q,4_Q)
+C_2^{0,X_2}(3_Q,\hat{1}_q;;5_q,6_g,\hat{2}_g,4_Q)
\nonumber\\
&&\hspace{1.0cm}
+\:C_2^{0,X_3}(3_Q,6_g,\hat{1}_q;;5_q,\hat{2}_g,4_Q)\Big)
+\tilde{C}_2^{0}(5_q,\hat{1}_q,3_Q,4_Q,\hat{2}_g,6_g)\nonumber\\
&&\hspace{1.0cm}
+\frac{1}{N_c^2}\tilde{\tilde{C}}_2^{0}(5_q,\hat{1}_q,3_Q,4_Q,\hat{2}_g,6_g)\Big)
\Big\}J^{(4)}_2(p_3,p_4,p_5,p_6),
\label{eq:nnloRRqg2}
\eeqa
where the leading-colour contribution consists of coherent squares of the leading-colour partial amplitudes while the 
subleading colour structures 
$\tilde{C}_2^0$ and $\tilde{\tilde{C}}_2^0$ are built from 
interferences between leading and subleading-colour amplitudes.
The overall normalisation factor for this channel is,
\beq
\norm_{\NNLO}^{\:qg}=\norm_{\NLO}^{\:qg}
\left(\frac{\alpha_s}{2\pi}\right)
\frac{\bar C(\epsilon)}{C(\epsilon)}.
\eeq

We now examine the unresolved limits of the leading-colour contribution, which determine the antenna structure required to describe its infrared behaviour. This contribution exhibits the following unresolved configurations:
\begin{itemize}
\item Triple-collinear limits:
\[
\hat{1}_q\parallel5_q\parallel6_g,\qquad
\hat{2}_g\parallel5_q\parallel6_g.
\]

\item Soft-collinear limits:
\[
6_g\to0+\hat{1}_q\parallel5_q,\qquad
6_g\to0+\hat{2}_g\parallel5_q.
\]

\item Double-collinear limits:
\[
\hat{1}_q\parallel5_q+\hat{2}_g\parallel6_g,\qquad
\hat{1}_q\parallel6_g+\hat{2}_g\parallel5_q.
\]

\item Single-unresolved limits:
\[
6_g\to0,\qquad
\hat{1}_q\parallel5_q,\qquad
\hat{2}_g\parallel5_q,\qquad
\hat{1}_q\parallel6_g,\qquad
\hat{2}_g\parallel6_g.
\]
\end{itemize}

The triple-collinear limit $\hat{1}_q\parallel5_q\parallel6_g$ can
be captured by the
initial-initial four-parton antennae
$G_4^0(\hat{2}_g,\hat{1}_q,5_q,6_g)$ and
$G_4^0(6_g,\hat{1}_q,5_q,\hat{2}_g)$,
which also reproduce the associated soft-collinear
$(6_g\to0+\hat{1}_q\parallel5_q)$
and double-collinear
$(\hat{1}_q\parallel5_q+\hat{2}_g\parallel6_g)$
limits.
Similarly, the initial-initial massless four-parton antennae
$A_4^0(\hat{1}_q,\hat{2}_g,6_g,5_q)$ and
$A_4^0(\hat{1}_q,6_g,\hat{2}_g,5_q)$
capture the
$\hat{2}_g\parallel5_q\parallel6_g$
triple-collinear limit together with the associated
soft-collinear
$(6_g\to0+\hat{2}_g\parallel5_q)$
and double-collinear
$(\hat{1}_q\parallel6_g+\hat{2}_g\parallel5_q)$
configurations.
The remaining single-unresolved limits are subtracted using three-parton antennae. 

In contrast to the purely fermionic channels, genuine single-soft gluon singularities are present. Since soft gluon emission depends on the adjacent colour-connected hard radiators, some of these limits involve a massive radiator and therefore require the use of massive antennae. 
The following expression for the leading-colour double-real subtraction term in the $qg$-channel reads:
\beqa
{\rm d}\hat{\sigma}_{qg,\NNLO}^{S,l.c.}&=&{\cal N}_{\NNLO}^{qg}\:
N_c^2\:{\rm d}\Phi_4(p_3,p_4,p_5,p_6;p_1,p_2)\Big\{\nonumber\\
\quad&\phantom{+}&D_{3}^{0}(\hat{1}_q,6_g,\hat{2}_g)\;C_1^{a}(\tilde{3}_Q,\hat{\bar{2}}_g,\hat{\bar{1}}_q;\tilde{5}_q,\tilde{4}_Q)\nonumber\\
\quad&+&d_3^0(4_Q,6_g,\hat{2}_g)\;C_1^{b}(3_Q,\hat{1}_q;5_q,\hat{\bar{2}}_g,(46)_Q)\nonumber\\
\quad&+&A_3^0(5_q,6_g,4_Q)\;C_1^{a}(3_Q,\hat{2}_g,\hat{1}_q;(56)_q,(46)_Q)\nonumber\\
&&\nonumber\\
&+&d_3^0(3_Q,6_g,\hat{2}_g)\;C_1^{a}((36)_Q,\hat{\bar{2}}_g,\hat{1}_q;5_q,4_Q)\nonumber\\
&+&d_3^0(5_q,6_g,\hat{2}_g)\;C_1^{b}(3_Q,\hat{1}_q;(56)_q,\hat{\bar{2}}_g,4_Q)\nonumber\\
&+&A_3^0(1_q,6_g,3_Q)\;C_1^{b}((36)_Q,\hat{\bar{1}}_q;5_q,\hat{2}_g,4_Q)\nonumber\\
&&\nonumber\\
&-&A_3^0(\hat{1}_q,\hat{2}_g,5_q)\;C_{1}^{a}(\tilde{3}_Q,\tilde{6}_g,\hat{1}_q;\hat{2}_q,\tilde{4}_Q)\nonumber\\
&-&A_3^0(\hat{1}_q,\hat{2}_g,5_q)\;C_{1}^{b}(\tilde{3}_Q,\hat{1}_q;\hat{2}_q,\tilde{6}_g,\tilde{4}_Q)\nonumber\\
&&\nonumber\\
&-&G_3^0(\hat{2}_g,\hat{1}_q,5_q)B_3^0(\tilde{3}_Q,\hat{\bar{1}}_g,\hat{\bar{2}}_g,\tilde{6}_g,\tilde{4}_Q)\nonumber\\
&-&G_3^0(\hat{2}_g,\hat{1}_q,5_q)B_3^0(\tilde{3}_Q,\hat{\bar{1}}_g,\tilde{6}_g,\hat{\bar{2}}_g,\tilde{4}_Q)\nonumber\\
&-&G_3^0(\hat{2}_g,\hat{1}_q,5_q)B_3^0(\tilde{3}_Q,\tilde{6}_g,\hat{\bar{1}}_g,\hat{\bar{2}}_g,\tilde{4}_Q)\nonumber\\
&-&G_3^0(\hat{2}_g,\hat{1}_q,5_q)B_3^0(\tilde{3}_Q,\hat{\bar{2}}_g,\hat{\bar{1}}_g,\tilde{6}_g,\tilde{4}_Q)\nonumber\\
&-&G_3^0(\hat{2}_g,\hat{1}_q,5_q)B_3^0(\tilde{3}_Q,\hat{\bar{2}}_g,\tilde{6}_g,\hat{\bar{1}}_g,\tilde{4}_Q)\nonumber\\
&-&G_3^0(\hat{2}_g,\hat{1}_q,5_q)B_3^0(\tilde{3}_Q,\tilde{6}_g,\hat{\bar{2}}_g,\hat{\bar{1}}_g,\tilde{4}_Q)\nonumber\\
&&\nonumber\\
&-&G_4^0(\hat{2}_g,\hat{1}_q,5_q,6_g)B_2^0(\tilde{3}_Q,\hat{\bar{1}}_g,\hat{\bar{2}}_g,\tilde{4}_Q)\nonumber\\
&+&d_3^0(5_q,6_g,\hat{2}_g)G_3^0(\hat{\bar{2}}_g,\hat{1}_q,(56)_q)B_2^0(\tilde{3}_Q,\hat{\bar{1}}_g,\hat{\bar{\bar{2}}}_g,\tilde{4}_Q)\nonumber\\
&+&G_3^0(\hat{2}_g,\hat{1}_q,5_q)F_3^0(\hat{\bar{1}}_g,\tilde{6}_g,\hat{\bar{2}}_g)B_2^0(\tilde{3}_Q,\hat{\bar{\bar{1}}}_g,\hat{\bar{\bar{2}}}_g,\tilde{4}_Q)\nonumber\\
&&\nonumber\\
&-&G_4^0(\hat{2}_g,\hat{1}_q,5_q,6_g)B_2^0(\tilde{3}_Q,\hat{\bar{2}}_g,\hat{\bar{1}}_g,\tilde{4}_Q)\nonumber\\
&+&d_3^0(5_q,6_g,\hat{2}_g)G_3^0(\hat{\bar{2}}_g,\hat{1}_q,(56)_q)B_2^0(\tilde{3}_Q,\hat{\bar{\bar{2}}}_g,\hat{\bar{1}}_g,\tilde{4}_Q)\nonumber\\
&+&G_3^0(\hat{2}_g,\hat{1}_q,5_q)F_3^0(\hat{\bar{1}}_g,\tilde{6}_g,\hat{\bar{2}}_g)B_2^0(\tilde{3}_Q,\hat{\bar{\bar{2}}}_g,\hat{\bar{\bar{1}}}_g,\tilde{4}_Q)\nonumber\\
&&\nonumber\\
&-&G_4^0(6_g,\hat{1}_q,5_q,\hat{2}_g)B_2^0(\tilde{3}_Q,\hat{\bar{1}}_g,\hat{\bar{2}}_g,\tilde{4}_Q)\nonumber\\
&+&D_3^0(\hat{1}_q,6_g,\hat{2}_g)G_3^0(\hat{\bar{2}}_g,\hat{\bar{1}}_q,\tilde{5}_q)B_2^0(\tilde{3}_Q,\hat{\bar{\bar{1}}}_g,\hat{\bar{\bar{2}}}_g,\tilde{4}_Q)\nonumber\\
&+&G_3^0(\hat{2}_g,\hat{1}_q,5_q)F_3^0(\hat{\bar{2}}_g,\tilde{6}_g,\hat{\bar{1}}_g)B_2^0(\tilde{3}_Q,\hat{\bar{\bar{1}}}_g,\hat{\bar{\bar{2}}}_g,\tilde{4}_Q)\nonumber\\
&+&A_3^0(\hat{1}_q,\hat{2}_g,5_q)G_3^0(\tilde{6}_g,\hat{\bar{2}}_q,\hat{\bar{1}}_q)B_2^0(\tilde{3}_Q,\hat{\bar{\bar{1}}}_g,\hat{\bar{\bar{2}}}_g,\tilde{4}_Q)\nonumber\\
&&\nonumber\\
&-&G_4^0(6_g,\hat{1}_q,5_q,\hat{2}_g)B_2^0(\tilde{3}_Q,\hat{\bar{2}}_g,\hat{\bar{1}}_g,\tilde{4}_Q)\nonumber\\
&+&D_3^0(\hat{1}_q,6_g,\hat{2}_g)G_3^0(\hat{\bar{2}}_g,\hat{\bar{1}}_q,\tilde{5}_q)B_2^0(\tilde{3}_Q,\hat{\bar{\bar{2}}}_g,\hat{\bar{\bar{1}}}_g,\tilde{4}_Q)\nonumber\\
&+&G_3^0(\hat{2}_g,\hat{1}_q,5_q)F_3^0(\hat{\bar{2}}_g,\tilde{6}_g,\hat{\bar{1}}_g)B_2^0(\tilde{3}_Q,\hat{\bar{\bar{2}}}_g,\hat{\bar{\bar{1}}}_g,\tilde{4}_Q)\nonumber\\
&+&A_3^0(\hat{1}_q,\hat{2}_g,5_q)G_3^0(\tilde{6}_g,\hat{\bar{2}}_q,\hat{\bar{1}}_q)B_2^0(\tilde{3}_Q,\hat{\bar{\bar{2}}}_g,\hat{\bar{\bar{1}}}_g,\tilde{4}_Q)\nonumber\\
&&\nonumber\\
&-&A_4^0(\hat{1}_q,\hat{2}_g,6_g,5_q)C_0^0(\hat{\bar{1}}_q,\tilde{3}_Q,\tilde{4}_Q,\hat{\bar{2}}_q)\nonumber\\
&+&d_3^0(5_q,6_g,\hat{2}_g)A_3^0(\hat{1}_q,\hat{\bar{2}}_g,(56)_q)C_0^0(\hat{\bar{1}}_q,\tilde{3}_Q,\tilde{4}_Q,\hat{\bar{\bar{2}}}_q)\nonumber\\
&&\nonumber\\
&-&A_4^0(\hat{1}_q,6_g,\hat{2}_g,5_q)C_0^0(\hat{\bar{1}}_q,\tilde{3}_Q,\tilde{4}_Q,\hat{\bar{2}}_q)\nonumber\\
&+&D_3^0(\hat{1}_q,6_g,\hat{2}_g)A_3^0(\hat{\bar{1}}_q,\hat{\bar{2}}_g,\tilde{5}_q)C_0^0(\hat{\bar{\bar{1}}}_q,\tilde{3}_Q,\tilde{4}_Q,\hat{\bar{\bar{2}}}_q)\nonumber\\
&+&A_3^0(\hat{1}_q,\hat{2}_g,5_q)A_3^0(\hat{\bar{1}}_q,\tilde{6}_g,\hat{\bar{2}}_q)C_0^0(\hat{\bar{\bar{1}}}_q,\tilde{3}_Q,\tilde{4}_Q,\hat{\bar{\bar{2}}}_q)\nonumber\\
&&\nonumber\\
&+&d_3^0(4_Q,6_g,\hat{2}_g)A_3^0(\hat{1}_q,\hat{\bar{2}}_g,5_q)C_0^0(\hat{\bar{1}}_q,3_Q,(46)_Q,\hat{\bar{\bar{2}}}_q)\nonumber\\
&+&A_3^0(\hat{1}_q,6_g,3_Q)A_3^0(\hat{\bar{1}}_q,\hat{2}_g,5_q)C_0^0(\hat{\bar{\bar{1}}}_q,\tilde{(36)}_Q,\tilde{4}_Q,\hat{\bar{2}}_q)\nonumber\\
&-&D_{3}^{0}(\hat{1}_q,6_g,\hat{2}_g)A_3^0(\hat{\bar{1}}_q,\hat{\bar{2}}_g,\tilde{5}_q)C_0^0(\hat{\bar{\bar{1}}}_q,\tilde{\tilde{3}}_Q,\tilde{\tilde{4}}_Q,\hat{\bar{\bar{2}}}_q)\nonumber\\
&&\nonumber\\
&+&d_3^0(4_Q,6_g,\hat{2}_g)G_3^0(\hat{\bar{2}}_g,\hat{1}_q,5_q)B_2^0(\tilde{3}_Q,\hat{\bar{1}}_g,\hat{\bar{\bar{2}}}_g,\tilde{(46)}_Q)\nonumber\\
&+&A_3^0(\hat{1}_q,6_g,3_Q)G_3^0(\hat{2}_g,\hat{\bar{1}}_q,5_q)B_2^0(\tilde{(36)}_Q,\hat{\bar{\bar{1}}}_g,\hat{\bar{2}}_g,\tilde{4}_Q)\nonumber\\
&-&D_{3}^{0}(\hat{1}_q,6_g,\hat{2}_g)G_3^0(\hat{\bar{2}}_g,\hat{\bar{1}}_q,\tilde{5}_q)B_2^0(\tilde{3}_Q,\hat{\bar{\bar{1}}}_g,\hat{\bar{\bar{2}}}_g,\tilde{4}_Q)\nonumber\\
&&\nonumber\\
&+&d_3^0(3_Q,6_g,\hat{2}_g)G_3^0(\hat{\bar{2}}_g,\hat{1}_q,5_q)B_2^0(\tilde{(36)}_Q,\hat{\bar{\bar{2}}}_g,\hat{\bar{1}}_g,\tilde{4}_Q)\nonumber\\
&+&A_3^0(5_q,6_g,4_Q)G_3^0(\hat{2}_g,\hat{1}_q,(56)_q)B_2^0(\tilde{3}_Q,\hat{\bar{2}}_g,\hat{\bar{1}}_g,\tilde{(46)}_Q)\nonumber\\
&-&d_3^0(5_q,6_g,\hat{2}_g)G_3^0(\hat{\bar{2}}_g,\hat{1}_q,(56)_q)B_2^0(\tilde{3}_Q,\hat{\bar{\bar{2}}}_g,\hat{\bar{1}}_g,\tilde{4}_Q)\nonumber\\
&&\nonumber\\
&+&(S_{\bar{\bar{1}}\tilde{6}\tilde{(36)}} + S_{\bar{2}\tilde{6}\tilde{4}} - S_{\bar{\bar{1}}\tilde{6}\bar{2}}-S_{264}-S_{\bar{1}6(36)}+S_{\bar{1}62})\nonumber\\
&&\hspace{0.3cm}\times
A_3^0(\hat{\bar{1}}_q,\hat{2}_g,5_q)C_0^0(\hat{\bar{\bar{1}}}_q,\tilde{(36)}_Q,\tilde{4}_Q,\hat{\bar{2}}_q)\nonumber\\
&&\nonumber\\
&+&(S_{\bar{2}\tilde{6}\tilde{4}}-S_{\bar{\bar{1}}\tilde{6}\tilde{(36)}} - S_{\bar{\bar{1}}\tilde{6}\bar{2}}-S_{264}-S_{\bar{1}6\tilde{(36)}}+S_{\bar{1}62})
\nonumber\\
&&\hspace{0.3cm}\times
G_3^0(\hat{2}_g,\hat{\bar{1}}_q,5_q)B_2^0(\tilde{(36)}_Q,\hat{\bar{\bar{1}}}_g,\hat{\bar{2}}_g,\tilde{4}_Q)\nonumber\\
&&\nonumber\\
&+&(S_{\bar{\bar{1}}\tilde{6}\tilde{4}}-S_{\bar{2}\tilde{6}\tilde{(36)}} - S_{\bar{\bar{1}}\tilde{6}\bar{2}}-S_{26(36)}-S_{564}+S_{265})
\nonumber\\
&&\hspace{0.3cm}\times
G_3^0(\hat{2}_g,\hat{\bar{1}}_q,5_q)B_2^0(\tilde{(36)}_Q,\hat{\bar{2}}_g,\hat{\bar{\bar{1}}}_g,\tilde{4}_Q)\Big\}\,.
\label{eq:qgsnnlo}
\eeqa
The terms proportional to $S_{ijk}$ in Eq.~\eqref{eq:qgsnnlo} 
constitute the large-angle soft contribution to the subtraction term 
${\rm d}\hat{\sigma}_{NNLO}^{S}$. Within the antenna-subtraction formalism, this contribution compensates the over-subtraction of the single-soft limit introduced by the iterated subtraction built from products of three-parton antennae~\cite{NigelGlover:2010kwr,Currie:2013vh,Chen:2022clm}. It is expressed in terms of eikonal factors whose hard radiators do not, in general, coincide with the radiators defining the antenna phase space.
For the present process, these eikonal factors 
involve either one massive and one massless hard radiator or two massless hard radiators. The analytic integration of these soft factors over the single-unresolved massive initial-final antenna phase space is presented here for the first time. Details of the calculation are given in Section~\ref{sec:massiveSF}.

\subsubsection{Real-virtual contribution in the $qg$-channel}
Unlike the purely fermionic channels discussed previously, the $qg$-channel receives a genuine real-virtual contribution arising from the interference of the one-loop and tree-level amplitudes for the partonic process $qg\to Q\bar{Q}q$. The colour decomposition for the
one-loop matrix element reads,
\begin{align}
&\mathcal{M}^{1}_{q_1 g_2 \to Q_3 \bar Q_4 q_5}
=
\sqrt{2}\:g_s^5\, \nonumber\\
&\hspace{0.7cm}\Big\{N_c
\Big[
(T^{a_2})_{i_3 i_1}\,\delta_{i_5 i_4}
\,\mathcal{M}^{1}_{5}(3_Q,\hat{2}_g,\hat{1}_{\bar{q}};;5_q,4_{\bar Q})
+(T^{a_2})_{i_5 i_4}\,\delta_{i_3 i_1}
\,\mathcal{M}^{1}_{5}(3_Q,\hat{1}_{\bar{q}};;\,5_q,\hat{2}_g,4_{\bar Q})\Big]
\nonumber \\
&\quad
-\frac{1}{N_c}\Big[
(T^{a_2})_{i_3 i_4}\,\delta_{i_5 i_1}
\,\mathcal{M}^{1}_{5}(3_Q,\hat{2}_g,4_{\bar Q};\,5_{q},\hat{1}_{\bar{q}})
+(T^{a_2})_{i_5 i_1}\,\delta_{i_3 i_4}
\,\mathcal{M}^{1}_{5}(3_Q,4_{\bar Q};\,5_q,\hat{2}_g,\hat{1}_{\bar{q}})
\Big]\Big\},
\label{eq:M1}
\end{align}
where each of the sub-amplitudes has the following decomposition
into primitive amplitudes
\begin{equation}
\mathcal{M}^{1}_{5}(\dots)
=
N_c\,\mathcal{M}^{[lc]}_{5}(\dots)
+ N_l\,\mathcal{M}^{[l]}_{5}(\dots)
+ N_h\,\mathcal{M}^{[h]}_{5}(\dots)
-
\frac{1}{N_c}\,\mathcal{M}^{[slc]}_{5}(\dots).
\end{equation}
Here, $N_l$, $N_h$ denote the numbers of light and heavy quark flavours, respectively, while
$\mathcal{M}^{[lc]}_{5}$,$\mathcal{M}^{[slc]}_{5}$,
$\mathcal{M}^{[l]}_{5}$,and $\mathcal{M}^{[h]}_{5}$ 
are the leading-colour, subleading-colour, light-fermion-loop and heavy-fermion-loop primitive amplitudes. 

Interfering the colour decomposition in Eq.~\eqref{eq:M1} with its tree-level counterpart gives the real-virtual contribution,
\begin{align}
&d\hat{\sigma}^{RV}_{q g,\,\text{NNLO}}=
{\cal N}_{\NNLO}^{qg}C(\epsilon)\int\:{\rm d}\Phi_3(p_3,p_4,p_5;p_1,p_2)
\:\Big\{\nonumber\\
&\hspace{1.0cm}
N_c^2
\;C_1^1(5_q,\hat{1}_{\bar{q}},3_Q,4_Q,\hat{2}_g)
+
\tilde{C}_1^1(5_q,\hat{1}_{\bar{q}},3_Q,4_Q,\hat{2}_g)
+\frac{1}{N_c^{2}}
\tilde{\tilde{C}}_1^1(5_q,\hat{1}_{\bar{q}},3_Q,4_Q,\hat{2}_g)
\nonumber \\
&\hspace{0.4cm}
+N_l \Big[ N_c
\;\hat{C}_1^{1,l}(5_q,\hat{1}_{\bar{q}},3_Q,4_Q,\hat{2}_g)
+\frac{1}{N_c}
\hat{\tilde{C}}_1^{1,l}(5_q,\hat{1}_{\bar{q}},3_Q,4_Q,\hat{2}_g)\Big]
\nonumber \\
&\quad
+ N_h \Big[ N_c\;
\hat{C}_1^{1,h}(5_q,\hat{1}_{\bar{q}},3_Q,4_Q,\hat{2}_g)
+\frac{1}{N_c}\;
\hat{\tilde{C}}_1^{1,h}(5_q,\hat{1}_{\bar{q}},3_Q,4_Q,\hat{2}_g)\Big]
\Big\}J^{(3)}_2(p_3,p_4,p_5).
\end{align}

Here the colour-ordered one-loop matrix elements are collected into
$C$-type matrix elements following the NNLOJET notation.
In total, seven independent colour structures contribute to the real-virtual matrix elements. Using the NNLOJET notation for colour-stripped matrix-elements, in the above equation a tilde indicates a subleading-colour contribution, while a hat identifies contributions proportional to closed light-$\hat{C}_1^{1,l}$ or heavy-fermion loops $\hat{C}_1^{1,h}$. 

The one-loop matrix elements are evaluated numerically using {\sc OpenLoops}~\cite{Buccioni:2019sur}. Consequently, the real-virtual contribution is implemented retaining its full colour dependence, as this is the level at which the one-loop amplitudes are provided by {\sc OpenLoops}. Accordingly, the explicit and implicit cancellation of infrared singularities is verified for the full-colour combination ${\rm d}\hat{\sigma}_{NNLO}^{RV}-{\rm d}\hat{\sigma}_{NNLO}^{T}$ 
of Eq.~\eqref{eq.subnnlo}. However, to keep the presentation compact, we display only the leading-colour subtraction terms in this section. 

For the leading colour part, besides the explicit infrared poles originating from the loop integration, the real-virtual matrix elements retain implicit singularities, related to initial-state collinear configurations. The relevant unresolved limits are $\hat{1}_q||5_q$, $\hat{2}_g||5_q$,
 while no soft singularities are present. The quark-quark collinear limits are described by the massless $G_3^{0}$ and $G_3^{1}$ 
 antennae, while the quark-gluon limits are captured by the corresponding
 $A_3^{0}$ and $A_3^{1}$ 
 antennae~\cite{Daleo:2006xa,Gehrmann:2011wi}. Together with the integrated double-real subtraction terms and the mass-factorisation counterterms, these contributions yield the following leading-colour expression for
 ${\rm d}\hat{\sigma}^{T}_{NNLO}$,
\beqa
{\rm d}\hat{\sigma}_{qg,\NNLO}^{T,l.c.}&=&{\cal N}_{\NNLO}^{qg}
C(\epsilon)\:N_c^2\:{\rm d}\Phi_3(p_3,p_4,p_5;p_1,p_2)\Big\{\nonumber\\&-&({\cal D}_{3,qg}^0(s_{12})-\Gamma_{qq}(x_1)-\frac{1}{2}\Gamma_{gg}(x_2))C_1^{0,a}(3_Q,\hat{2}_g,\hat{1}_q;5_q,4_Q)\nonumber\\
&-&({\cal D}_{3,g}^{0m}(s_{24})-\frac{1}{2}\Gamma_{gg}(x_2))C_1^{0,b}(3_Q,\hat{1}_q;5_q,\hat{2}_g,4_Q)\nonumber\\
&-&{\cal A}_3^{0m}(s_{45})C_1^{0,a}(3_Q,\hat{2}_g,\hat{1}_q;5_q,4_Q)\nonumber\\
&&\nonumber\\
&-&({\cal D}_{3,g}^{0m}(s_{23})-\frac{1}{2}\Gamma_{gg}(x_2))C_1^{0,a}(3_Q,\hat{2}_g,\hat{1}_q;5_q,4_Q)\nonumber\\
&-&({\cal D}_{3,g}^{0}(s_{25})-\frac{1}{2}\Gamma_{gg}(x_2))C_1^{0,b}(3_Q,\hat{1}_q;5_q,\hat{2}_g,4_Q)\nonumber\\
&-&({\cal A}_{3,q}^{0m}(s_{13})-\Gamma_{qq}(x_1))C_1^{0,b}(3_Q,\hat{1}_q;5_q,\hat{2}_g,4_Q)\nonumber\\
&&\nonumber\\
&+&({\cal A}_{3,qg}^0(s_{12})+S_{g\to q}\Gamma_{qg}(x_2))C_1^{0,a}(3_Q,5_g,\hat{1}_q;\hat{2}_q,4_Q)\nonumber\\
&+&({\cal A}_{3,qg}^0(s_{12})+S_{g\to q}\Gamma_{qg}(x_2))C_1^{0,b}(3_Q,\hat{1}_q;\hat{2}_q,5_g,4_Q)\nonumber\\
&&\nonumber\\
&-&({\cal G}_{3,qg}^0(s_{12})+S_{q\to g}\Gamma_{gq}(x_1))B_3^0(3_Q,\hat{1}_g,\hat{2}_g,5_g,4_Q)\nonumber\\
&-&({\cal G}_{3,qg}^0(s_{12})+S_{q\to g}\Gamma_{gq}(x_1))B_3^0(3_Q,\hat{1}_g,5_g,\hat{2}_g,4_Q)\nonumber\\
&-&({\cal G}_{3,qg}^0(s_{12})+S_{q\to g}\Gamma_{gq}(x_1))B_3^0(3_Q,5_g,\hat{1}_g,\hat{2}_g,4_Q)\nonumber\\
&-&({\cal G}_{3,qg}^0(s_{12})+S_{q\to g}\Gamma_{gq}(x_1))B_3^0(3_Q,\hat{2}_g,\hat{1}_g,5_g,4_Q)\nonumber\\
&-&({\cal G}_{3,qg}^0(s_{12})+S_{q\to g}\Gamma_{gq}(x_1))B_3^0(3_Q,\hat{2}_g,5_g,\hat{1}_g,4_Q)\nonumber\\
&-&({\cal G}_{3,qg}^0(s_{12})+S_{q\to g}\Gamma_{gq}(x_1))B_3^0(3_Q,5_g,\hat{2}_g,\hat{1}_g,4_Q)\nonumber\\
&&\nonumber\\
&-&G_3^0(\hat{2}_g,\hat{1}_q,5_q)B_2^1(\tilde{3}_Q,\hat{\bar{1}}_g,\hat{\bar{2}}_g,\tilde{4}_Q)\nonumber\\
&-&G_3^0(\hat{2}_g,\hat{1}_q,5_q)({\cal D}_{3,g}^{0m}(s_{\bar{1}\tilde{3}}) + {\cal F}_3^0(s_{\bar{1}\bar{2}})
+{\cal D}_{3,g}^{0m}(s_{\bar{2}\tilde{4}})\nonumber\\
&&\hspace{1.5cm}-\Gamma_{gg}(x_1)-\Gamma_{gg}(x_2) )B_2^0(\tilde{3}_Q,\hat{\bar{1}}_g,\hat{\bar{2}}_g,\tilde{4}_Q)\nonumber\\
&-&G_3^0(\hat{2}_g,\hat{1}_q,5_q)B_2^1(\tilde{3}_Q,\hat{\bar{2}}_g,\hat{\bar{1}}_g,\tilde{4}_Q)\nonumber\\
&-&G_3^0(\hat{2}_g,\hat{1}_q,5_q)({\cal D}_{3,g}^{0m}(s_{\bar{2}\tilde{3}}) + {\cal F}_3^0(s_{\bar{1}\bar{2}})
+{\cal D}_{3,g}^{0m}(s_{\bar{1}\tilde{4}})\nonumber\\
&&\hspace{1.5cm}-\Gamma_{gg}(x_1)-\Gamma_{gg}(x_2) )B_2^0(\tilde{3}_Q,\hat{\bar{2}}_g,\hat{\bar{1}}_g,\tilde{4}_Q)\nonumber\\
&&\nonumber\\
&-&G_3^1(\hat{2}_g,\hat{1}_q,5_q)B_2^0(\tilde{3}_Q,\hat{\bar{1}}_g,\hat{\bar{2}}_g,\tilde{4}_Q)\nonumber\\
&-&G_3^0(\hat{2}_g,\hat{1}_q,5_q)({\cal D}_{3,g}^{0}(s_{25})+{\cal D}_{3,qg}^{0}(s_{12})- 2{\cal F}_3^0(s_{\bar{1}\bar{2}})\nonumber\\
&&\hspace{1.5cm}
-\Gamma_{qq}(x_1)+\Gamma_{gg}(x_1))B_2^0(\tilde{3}_Q,\hat{\bar{1}}_g,\hat{\bar{2}}_g,\tilde{4}_Q)\nonumber\\
&-&G_3^1(\hat{2}_g,\hat{1}_q,5_q)B_2^0(\tilde{3}_Q,\hat{\bar{2}}_g,\hat{\bar{1}}_g,\tilde{4}_Q)\nonumber\\
&-&G_3^0(\hat{2}_g,\hat{1}_q,5_q)({\cal D}_{3,g}^{0}(s_{25})+{\cal D}_{3,qg}^{0}(s_{12}) - 2{\cal F}_3^0(s_{\bar{1}\bar{2}})\nonumber\\
&&\hspace{1.5cm}
-\Gamma_{qq}(x_1)+\Gamma_{gg}(x_1))B_2^0(\tilde{3}_Q,\hat{\bar{2}}_g,\hat{\bar{1}}_g,\tilde{4}_Q)\nonumber\\
&&\nonumber\\
&-&A_3^0(5_q,\hat{2}_g,\hat{1}_q)C_0^1(\hat{\bar{1}}_q,\tilde{3}_Q,\tilde{4}_Q,\hat{\bar{2}}_q)\nonumber\\
&-&A_3^0(5_q,\hat{2}_g,\hat{1}_q)({\cal A}_q^{0m}(s_{\bar{1}\tilde{3}})+{\cal A}_q^{0m}(s_{\bar{2}\tilde{4}})\nonumber\\
&&\hspace{1.5cm}-\Gamma_{qq}(x_1)-\Gamma_{qq}(x_2))C_0^0(\hat{\bar{1}}_q,\tilde{3}_Q,\tilde{4}_Q,\hat{\bar{2}}_q)\nonumber\\
&&\nonumber\\
&-&A_3^1(5_q,\hat{2}_g,\hat{1}_q)C_0^0(\hat{\bar{1}}_q,\tilde{3}_Q,\tilde{4}_Q,\hat{\bar{2}}_q)\nonumber\\
&-&A_3^0(5_q,\hat{2}_g,\hat{1}_q)({\cal D}_g^{0}(s_{25})+{\cal D}_{3,qg}^{0}(s_{12})-{\cal A}_{3,qq}^{0}(s_{\bar{1}\bar{2}})\nonumber\\
&&\hspace{1.5cm}+\Gamma_{qq}(x_2)-\Gamma_{gg}(x_2))C_0^0(\hat{\bar{1}}_q,\tilde{3}_Q,\tilde{4}_Q,\hat{\bar{2}}_q)\nonumber\\
&&\nonumber\\
&-&A_3^0(\hat{1}_q,5_g,3_Q)({\cal A}_{3,qg}^0(s_{\bar{1}2})+S_{g\to q}\Gamma_{qg}(x_2))C_0^0(\hat{\bar{1}}_q,(35)_Q,4_Q,\hat{2}_q)\nonumber\\
&-&A_3^0(\hat{2}_q,5_g,4_Q)({\cal A}_{3,qg}^0(s_{1\bar{2}})+S_{g\to q}\Gamma_{qg}(x_2))C_0^0(\hat{1}_q,3_Q,(45)_Q,\hat{\bar{2}}_q)\nonumber\\
&&\nonumber\\
&-&d_3^0(4_Q,5_g,\hat{2}_g)({\cal G}_{3,qg}^0(s_{1\bar{2}})+S_{q\to g}\Gamma_{gq}(x_1))B_2^0(3_Q,\hat{1}_g,\hat{\bar{2}}_g,(45)_Q)\nonumber\\
&-&F_3^0(\hat{1}_g,5_g,\hat{2}_g)({\cal G}_{3,qg}^0(s_{\bar{1}\bar{2}})+S_{q\to g}\Gamma_{gq}(x_1))B_2^0(\tilde{3}_Q,\hat{\bar{1}}_g,\hat{\bar{2}}_g,\tilde{4}_Q)\nonumber\\
&-&d_3^0(3_Q,5_g,\hat{1}_g)({\cal G}_{3,qg}^0(s_{\bar{1}2})+S_{q\to g}\Gamma_{gq}(x_1))B_2^0((35)_Q,\hat{\bar{1}}_g,\hat{2}_g,4_Q)\nonumber\\
&-&d_3^0(4_Q,5_g,\hat{1}_g)({\cal G}_{3,qg}^0(s_{\bar{1}2})+S_{q\to g}\Gamma_{gq}(x_1))B_2^0(3_Q,\hat{2}_g,\hat{\bar{1}}_g,(45)_Q)\nonumber\\
&-&F_3^0(\hat{1}_g,5_g,\hat{2}_g)({\cal G}_{3,qg}^0(s_{\bar{1}\bar{2}})+S_{q\to g}\Gamma_{gq}(x_1))B_2^0(\tilde{3}_Q,\hat{\bar{2}}_g,\hat{\bar{1}}_g,\tilde{4}_Q)\nonumber\\
&-&d_3^0(3_Q,5_g,\hat{2}_g)({\cal G}_{3,qg}^0(s_{1\bar{2}})+S_{q\to g}\Gamma_{gq}(x_1))B_2^0((35)_Q,\hat{\bar{2}}_g,\hat{1}_g,4_Q)\nonumber\\
&&\nonumber\\
&-&2F_3^0(\hat{1}_g,5_g,\hat{2}_g)({\cal G}_{3,qg}^0(s_{12})-{\cal G}_{3,qg}^0(s_{\bar{1}\bar{2}}))B_2^0(\tilde{3}_Q,\hat{\bar{1}}_g,\hat{\bar{2}}_g,\tilde{4}_Q)\nonumber\\
&-&2F_3^0(\hat{1}_g,5_g,\hat{2}_g)({\cal G}_{3,qg}^0(s_{12})-{\cal G}_{3,qg}^0(s_{\bar{1}\bar{2}}))B_2^0(\tilde{3}_Q,\hat{\bar{2}}_g,\hat{\bar{1}}_g,\tilde{4}_Q)\nonumber\\
&&\nonumber\\
&-&G_3^0(5_g,\hat{1}_q,\hat{2}_q)({\cal A}_{3,qg}^0(s_{12})-{\cal A}_{3,qg}^0(s_{\bar{1}\bar{2}}))B_2^0(\tilde{3}_Q,\hat{\bar{1}}_g,\hat{\bar{2}}_g,\tilde{4}_Q)\nonumber\\
&-&G_3^0(5_g,\hat{1}_q,\hat{2}_q)({\cal A}_{3,qg}^0(s_{12})-{\cal A}_{3,qg}^0(s_{\bar{1}\bar{2}}))B_2^0(\tilde{3}_Q,\hat{\bar{2}}_g,\hat{\bar{1}}_g,\tilde{4}_Q)\nonumber\\
&&\nonumber\\
&-&A_3^0(\hat{1}_q,5_g,\hat{2}_q)({\cal A}_{3,qg}^0(s_{12})-{\cal A}_{3,qg}^0(s_{\bar{1}\bar{2}}))C_0^0(\hat{\bar{1}}_q,\tilde{3}_Q,\tilde{4}_Q,\hat{\bar{2}}_q)\nonumber\\
&&\nonumber\\
&+&A_3^0(5_q,\hat{2}_g,\hat{1}_q)C_0^0(\hat{\bar{1}}_q,\tilde{3}_Q,\tilde{4}_Q,\hat{\bar{2}}_q)\Big(\nonumber\\
&&-{\cal D}_{3,g}^{0m}(s_{24})-{\cal A}_{3,q}^{0m}(s_{13})+{\cal D}_{3,qg}^{0}(s_{12})+{\cal A}_{3,q}^{0m}(s_{\bar{2}\tilde{4}})+{\cal A}_{3,q}^{0m}(s_{\bar{1}\tilde{3}})-{\cal A}_{3,qq}^{0}(s_{\bar{1}\bar{2}})\nonumber\\
&&-\tilde{\cal S}_{0m}^{0m}(s_{\bar{1}\tilde{3}},s_{\bar{1}\tilde{3}})-\tilde{\cal S}_{0m}^{0m}(s_{\bar{2}\tilde{4}},s_{\bar{1}\tilde{3}})
+\tilde{\cal S}_{00}^{0m}(s_{\bar{1}\bar{2}},s_{\bar{1}\tilde{3}})
+{\cal S}_{0m}^{0m}(s_{24},s_{13})\nonumber\\
&&+{\cal S}_{0m}^{0m}(s_{13},s_{13})-{\cal S}_{00}^{0m}(s_{12},s_{13})\Big)\nonumber\\
&&\nonumber\\
&+&G_3^0(\hat{2}_g,\hat{1}_q,5_q)B_2^0(\tilde{3}_Q,\hat{\bar{1}}_g,\hat{\bar{2}}_g,\tilde{4}_Q)\Big(\nonumber\\
&&-{\cal D}_{3,g}^{0m}(s_{24})-{\cal A}_{3,q}^{0m}(s_{13})+{\cal D}_{3,qg}^{0}(s_{12})+{\cal D}_{3,g}(s_{\bar{2}\tilde{4}})+{\cal D}_{3,g}(s_{\bar{1}\tilde{3}})-{\cal F}_{3}^0(s_{\bar{1}\bar{2}})\nonumber\\
&&-\tilde{\cal S}_{0m}^{0m}(s_{\bar{1}\tilde{3}},s_{\bar{1}\tilde{3}})-\tilde{\cal S}_{0m}^{0m}(s_{\bar{2}\tilde{4}},s_{\bar{1}\tilde{3}})
+\tilde{\cal S}_{00}^{0m}(s_{\bar{1}\bar{2}},s_{\bar{1}\tilde{3}})
+{\cal S}_{0m}^{0m}(s_{24},s_{13})\nonumber\\
&&+{\cal S}_{0m}^{0m}(s_{13},s_{13})-{\cal S}_{00}^{0m}(s_{12},s_{13})\Big)\nonumber\\
&&\nonumber\\
&+&G_3^0(\hat{2}_g,\hat{1}_q,5_q)B_2^0(\tilde{3}_Q,\hat{\bar{2}}_g,\hat{\bar{1}}_g,\tilde{4}_Q)\Big(\nonumber\\
&&-{\cal D}_{3,g}^{0m}(s_{23})-{\cal A}_3^{0m}(s_{45})+{\cal D}_{3,g}^{0}(s_{25})+{\cal D}_{3,g}^{0m}(s_{\bar{2}\tilde{3}})+{\cal D}_{3,g}^{0m}(s_{\bar{1}\tilde{4}})-{\cal F}_{3}^0(s_{\bar{1}\bar{2}})\nonumber\\
&&-\tilde{\cal S}_{0m}^{0m}(s_{\bar{1}\tilde{4}},s_{\bar{1}\tilde{3}})-\tilde{\cal S}_{0m}^{0m}(s_{\bar{2}\tilde{3}},s_{\bar{1}\tilde{3}})+\tilde{\cal S}_{00}^{0m}(s_{\bar{1}\bar{2}},s_{\bar{1}\tilde{3}})
+{\cal S}_{0m}^{0m}(s_{23},s_{13})\nonumber\\
&&+{\cal S}_{0m}^{0m}(s_{45},s_{13})-{\cal S}_{00}^{0m}(s_{25},s_{13})\Big)\Big\}.
\label{eq:qgtnnlo}
\eeqa
The calligraphic symbols ${\cal S}$ denote the integrated soft factors associated with the large-angle soft subtraction term
present in Eq.~\eqref{eq:qgsnnlo}.
They provide the integrated counterpart of ${\rm d}\hat{\sigma}_{NNLO}^{S}$ 
and are required for the analytic cancellation of the explicit infrared poles of the real-virtual contribution. Their derivation is presented in Section~\ref{sec:massiveSF}.

\subsubsection{Double-virtual contribution in the $qg$-channel}
The final ingredient of the NNLO correction in the 
$qg$-channel is the contribution ${\rm d}\hat{\sigma}^{U}_{\NNLO}$.
While no genuine double-virtual matrix element is present in this
partonic channel, this contribution
combines all building blocks developed throughout this work. 
Upon integration over the unresolved phase space, the singular limits of the real-real and real-virtual 
radiation contributions factorise onto
matrix elements corresponding to either the
$q\bar{q}\to Q\bar{Q}$ or $gg\to Q\bar{Q}$ channels.
Accordingly, the ${\rm d}\hat{\sigma}^{U}_{\NNLO}$ contribution
naturally decomposes into terms proportional to
$q\bar{q}$ initiated ($C$-type) and $gg$ initiated ($B$-type) matrix elements.
These contributions can be organised in terms of integrated identity-changing dipoles, comprising integrated antenna functions together with the associated mass-factorisation kernels. Restricting ourselves to the leading-colour contribution, the resulting expression is,
\beqa
{\rm d}\hat{\sigma}_{qg,\NNLO}^{U,l.c.}&=&
{\cal N}_{\NNLO}^{qg}C(\epsilon)^2\:
N_c^2\:{\rm d}\Phi_3(p_3,p_4;p_1,p_2)\Big\{\nonumber\\
\quad&-&\mathcal{J}_{2,g\to q}^{(1)}(\hat{1}_q,\hat{2}_{g\to q})\;C_{0}^{1}(\hat{1}_q,3_Q,4_Q,\hat{2}_q)\nonumber\\
\quad&-&\mathcal{J}_{2,g\to q}^{(1)}(\hat{1}_q,\hat{2}_{g\to q})\otimes \mathcal{J}_{4}^{1}(\hat{1}_q,3_Q,4_Q,\hat{2}_q)\;C_{0}^{0}(\hat{1}_q,3_Q,4_Q,\hat{2}_q)\nonumber\\
\quad&-&\mathcal{J}_{2,g\to q}^{(2)}(\hat{1}_q,\hat{2}_{g\to q})\;C_{0}^{0}(\hat{1}_q,3_Q,4_Q,\hat{2}_q)\nonumber\\
\nonumber\\
\quad&-&\mathcal{J}_{2,q\to g}^{(1)}(\hat{1}_{q\to g},\hat{2}_{g})\;B_{2}^{1}(3_Q,\hat{1}_g,\hat{2}_g,4_Q)\nonumber\\
\quad&-&\mathcal{J}_{2,q\to g}^{(1)}(\hat{1}_{q\to g},\hat{2}_{g})\;B_{2}^{1}(3_Q,\hat{2}_g,\hat{1}_g,4_Q)\nonumber\\
\quad&-&\mathcal{J}_{2,q\to g}^{(1)}(\hat{1}_{q\to g},\hat{2}_{g})\otimes \mathcal{J}_{4}^{1}(3_Q,\hat{1}_g,\hat{2}_g,4_Q) \;B_{2}^{0}(3_Q,\hat{1}_g,\hat{2}_g,4_Q)\nonumber\\
\quad&-&\mathcal{J}_{2,q\to g}^{(1)}(\hat{1}_{q\to g},\hat{2}_{g})\otimes \mathcal{J}_{4}^{1}(3_Q,\hat{2}_g,\hat{1}_g,4_Q) \;B_{2}^{0}(3_Q,\hat{2}_g,\hat{1}_g,4_Q)\nonumber\\
\quad&-&\mathcal{J}_{2,q\to g}^{(2)}(\hat{1}_{q\to g},\hat{2}_{g})\;B_{2}^{0}(3_Q,\hat{1}_g,\hat{2}_g,4_Q)\quad\nonumber\\
\quad&-&\mathcal{J}_{2,q\to g}^{(2)}(\hat{1}_{q\to g},\hat{2}_{g})\;B_{2}^{0}(3_Q,\hat{2}_g,\hat{1}_g,4_Q)\Big\}\,.
\label{eq:Uqg}
\eeqa

The one-loop integrated operators $\mathcal{J}_{4}^{1}$ appearing above encode the explicit infrared pole structure of the reduced one-loop matrix elements $C_1^1$ and $B_2^1$ and are given by the sum of the corresponding colour-connected integrated dipoles,
\begin{eqnarray}
\mathcal{J}_{4}^{1}(\hat{1}_q,3_Q,4_Q,\hat{2}_q)
&=&
\mathcal{J}_{2}^{(1)}(\hat{1}_q,3_Q)
+\mathcal{J}_{2}^{(1)}(\hat{2}_q,4_Q)\\
\mathcal{J}_{4}^{1}(3_Q,\hat{1}_g,\hat{2}_g,4_Q)
&=&
\mathcal{J}_{2}^{(1)}(\hat{1}_g,3_Q)
+\mathcal{J}_{2}^{(1)}(\hat{1}_g,\hat{2}_g)
+\mathcal{J}_{2}^{(1)}(\hat{2}_g,4_Q).
\end{eqnarray}
The individual integrated dipoles are,
\begin{eqnarray}
\mathcal{J}_2^{(1)}(\hat{1}_q,3_Q)
&=&
{\cal A}_{3,q}^{0,m}(s_{13})
-\Gamma^{(1)}_{qq}(x_1)
\nonumber\\
\mathcal{J}_2^{(1)}(\hat{2}_{\bar q},4_{\bar Q})
&=&
{\cal A}_{3,q}^{0,m}(s_{24})
-\Gamma^{(1)}_{qq}(x_2)
\nonumber\\
\mathcal{J}_2^{(1)}(\hat{1}_g,3_Q)
&=&
{\cal D}_{3,g}^{0,m}(s_{13})
-\frac{1}{2}\Gamma^{(1)}_{gg}(x_1)
\nonumber\\
\mathcal{J}_2^{(1)}(\hat{2}_g,4_{\bar Q})
&=&
{\cal D}_{3,g}^{0,m}(s_{24})
-\frac{1}{2}\Gamma^{(1)}_{gg}(x_2)
\nonumber\\
\mathcal{J}_2^{(1)}(\hat{1}_g,\hat{2}_g)
&=&
{\cal F}_{3,gg}^{0}(s_{12})
-\frac{1}{2}\Gamma^{(1)}_{gg}(x_1)
-\frac{1}{2}\Gamma^{(1)}_{gg}(x_2).
\end{eqnarray}

The remaining single integrated operators are the identity-changing dipoles associated with the initial-state flavour transitions
$g\to q$ and $q\to g$. They combine the corresponding integrated three-parton antennae with the first-order mass-factorisation kernels and govern the factorisation of the unresolved contributions onto the
$q\bar{q}$ and $gg$ channels, respectively. They are defined by,
\beqa
\mathcal{J}_{2,g\to q}^{(1)}(\hat{1}_{q},\hat{2}_{g\to q})
&=&
-{\cal A}_{3,qg}^{0}(s_{12})
-S_{g\to q}\Gamma^{(1)}_{qg}(x_2)
\nonumber\\
\mathcal{J}_{2,q\to g}^{(1)}(\hat{1}_{q\to g},\hat{2}_{g})
&=&
-{\cal G}_{3,qg}^{0}(s_{12})
-S_{q\to g}\Gamma^{(1)}_{gq}(x_1).
\eeqa
The corresponding double integrated identity-changing dipoles
$\mathcal{J}_{2,g\to q}^{(2)}(\hat{1}_{q},\hat{2}_{g\to q})$
and\\ 
$\mathcal{J}_{2,q\to g}^{(2)}(\hat{1}_{q\to g},\hat{2}_{g})$
arise naturally in the present calculation. These massless operators are, already known from previous calculations performed within the NNLOJET framework~\cite{Chen:2022clm}, 
providing a further consistency check of the integrated subtraction framework. They are given by,
\beqa
\mathcal{J}_{2,g\to q}^{(2)}(\hat{1}_{q},\hat{2}_{g\to q})&=&-{\cal A}_{4,qg}^0(s_{12})-{\cal A}_{4,qg}^{0,n.adj}(s_{12})-{\cal A}_{3,qg}^1(s_{12})-\frac{b_0}{\epsilon}{\cal A}_{3,qg}^0(s_{12})\left(\frac{s_{12}}{\mu^2}\right)^{-\epsilon}\nonumber\\
&+&\frac{b_0}{\epsilon}{\cal A}_{3,qg}^0(s_{12})+[ {\cal A}_{3,qg}^0\otimes {\cal A}_{3,qg}^0] (s_{12})-[\Gamma^{1}_{qq,2} \otimes {\cal A}_{3,qg}^0] (s_{12})\nonumber\\
&+&[\Gamma^{1}_{gg,2} \otimes {\cal A}_{3,qg}^0] (s_{12})
-\frac{1}{2}S_{g\to q}[\Gamma^{1}_{qg,2}\otimes \Gamma^{1}_{qq,2}]\nonumber\\
&+&\frac{1}{2}S_{g\to q}[\Gamma^{1}_{qg,2}\otimes \Gamma^{1}_{gg,2}]-S_{g\to q}\bar{\Gamma}_{qg,2}^{(2)}(x_2)+\frac{b_0}{\epsilon}S_{g\to q}\Gamma^{(1)}_{qg}(x_2)\,,\\
\mathcal{J}_{2,q\to g}^{(2)}(\hat{1}_{q\to g},\hat{2}_{g})&=&-{\cal G}_{4,qg}^0(s_{12})-{\cal G}_{4,qg}^{0,n.adj}(s_{12})-{\cal G}_{3,qg}^{1}(s_{12})-\frac{b_0}{\epsilon}{\cal G}_{3,qg}^0(s_{12})\left(\frac{s_{12}}{\mu^2}\right)^{-\epsilon}\nonumber\\
&+&\frac{b_0}{\epsilon}{\cal G}_{3,qg}^0(s_{12})+2[{\cal G}_{3,qg}^0\otimes {\cal F}_{3,gg}^0]-S_{q\to g}\bar{\Gamma}^{(2)}_{gq}(x_1)+\frac{b_0}{\epsilon}S_{q\to g}\Gamma^{(1)}_{gq}(x_1)\nonumber\\
&+&[{\cal A}_{3,qg}^{0}\otimes {\cal G}_{3,qq}^{0}]+[\Gamma_{qq,1}^{(0)}\otimes {\cal G}_{3,qg}^{0}]-[\Gamma_{gg,1}^{(0)}\otimes {\cal G}_{3,qg}^{0}]\nonumber\\
&+&\frac{1}{2}S_{q\to g}[\Gamma_{qq,1}^{(0)}\otimes \Gamma_{gq,1}^{(0)}]-\frac{1}{2}S_{q\to g}[\Gamma_{gg,1}^{(0)}\otimes \Gamma_{gq,1}^{(0)}]\,.
\eeqa
Equation~\eqref{eq:Uqg} brings together all integrated building blocks required for the double virtual contribution in the $qg$-channel.
In particular, the products of integrated dipoles involve new convolution integrals between integrated massive and massless three-parton antennae, together with the corresponding mass-factorisation kernels. These integrals were evaluated analytically in the present work, extending the integrated antenna library required for NNLO calculations involving massive fermions.
These results complete the description of the subtraction terms required at ${\cal O}(\alpha_s^4)$ for the $qg$-channel.  
We further verified that all explicit infrared poles cancel analytically, providing a stringent validation of the entire framework. The corresponding numerical validation of the complete implementation is presented in Section~\ref{sec:nnlonumerics}.

\section{Analytic integration of soft factors over the initial-final massive antenna phase space}
\label{sec:massiveSF}

In addition to subtraction terms associated with unresolved radiation between pairs of colour-connected hard radiators, the antenna-subtraction formalism requires dedicated subtraction terms to account for large-angle soft gluon emission in so-called almost-colour connected configurations ~\cite{NigelGlover:2010kwr,Currie:2013vh,Chen:2022clm}. These contributions compensate the over-subtraction of the single-soft limit introduced by the iterated subtraction built from products of three-parton antennae. They therefore constitute an essential part of the subtraction procedure. Large-angle soft subtraction terms have previously been derived and integrated over massless antenna phase spaces within the antenna-subtraction formalism~\cite{Gehrmann-DeRidder:2007foh,Daleo:2009yj}. The $qg$-channel studied here provides the first NNLO application to top-quark pair production that requires these contributions within the massive antenna-subtraction formalism. 

In the following, we shall define the soft factor contributions, present 
in $\rm{d}\hat{\sigma}_{NNLO}^{S}$ in Eq.~\eqref{eq:qgsnnlo} and derive
the corresponding integrated soft factors contributing
to $\rm{d}\hat{\sigma}_{NNLO}^{T}$ in Eq.~\eqref{eq:qgtnnlo}.
For the quark-gluon channel, two distinct classes of soft factors arise. The first corresponds to a soft gluon emitted between a massive and a massless hard radiator, while the second involves two massless hard radiators. In both cases, the soft factors are integrated over an initial-final massive antenna phase space with one massive radiator. The derivation of these two classes of integrated soft factors is presented in 
Sections~\ref{sec:SFint1} and~\ref{sec:SFint2}, respectively. 
In addition, in Section~\ref{sec:SFnumcheck} we present a dedicated numerical consistency check that validates both the integrated soft factors derived in this section and their implementation in NNLOJET.

\subsection{Massive soft factor integrated over the initial-final massive  phase space}
\label{sec:SFint1}
The large-angle soft terms are obtained from eikonal factors involving two hard spectator partons. In the present case, we consider a massless spectator 
with momentum ($p_a$) and a massive spectator with momentum ($p_c$) and mass $p_c^2=m^2$, together with the unresolved gluon of momentum ($p_j$). The massive unintegrated soft factor is given by
\begin{equation}
S_{ajc}^{0m}=\frac{2s_{ac}}{s_{aj}s_{jc}}-\frac{2m^2}{s_{jc}^2},
\label{eq:SF0m}
\end{equation}
where $(s_{rs}=2p_r\cdot p_s).$

The integrated counterpart of~\eqref{eq:SF0m} is required over the initial-final 
massive antenna phase space. Following the standard antenna formalism~\cite{Abelof:2011jv}, the initial-final massive antenna phase space is defined through the two-particle 
process $q+p_i\rightarrow p_j+p_k$, where $q$ denotes the momentum carried by the off-shell current, satisfying $Q^2=-q^2>0$. 
The momentum $p_i$ is the initial-state massless radiator, $p_j$ is the unresolved soft gluon, and $p_k$ is the final-state massive radiator 
with $p_i^2=p_j^2=0, p_k^2=m^2$. The mass of the final-state radiator is taken to be equal to that of the massive spectator appearing in the eikonal factor, $p_k^2=p_c^2=m^2$.
The initial-final massive antenna phase space is 
conveniently expressed in terms of the two-particle phase space~\cite{Abelof:2011jv},
\begin{equation}
{\rm d}X_{i,jk}(p_j,p_k;p_i,q)=\frac{Q^2+m^2}{2\pi}\;{\rm d}\Phi_2(p_j,p_k;p_i,q),
\label{eq:IFantPS}
\end{equation}
with $q=p_j+p_k-p_i$ and $Q^2=-q^2>0$. 

It is convenient to parametrize the unresolved momentum $p_j$ in  spherical coordinates in $d-1$ spatial dimensions, where the radial coordinate 
is identified with the gluon energy and the remaining coordinates correspond to the angular variables. 
In this parametrization, the $(d-1)$-dimensional momentum-space measure factorizes into an energy-dependent part and an angular part according to
\begin{equation}
{\rm d}^{d-1}p_j=E_j^{d-2}{\rm d}E_j\;(\sin\theta_j)^{d-4}{\rm d}(\cos\theta_j)\;(\sin\psi_j)^{d-5}{\rm d}(\cos\psi_j){\rm d}\Omega_{d-4},
\end{equation}
where ${\rm d}\Omega_{d-4}$ is the differential solid angle on the $(d-4)$-dimensional unit sphere. 

In terms of the energy and angular variables of the unresolved momentum $p_j$, the phase space is
\begin{eqnarray} 
{\rm d}X_{i,jk}(p_j,p_k;p_i,q)&=&\left(\frac{Q^2+m^2}{2\pi}\right)\,\frac{(2\pi)^{2-d}}{4E_{cm}}\,E_j^{d-3}{\rm d}E_j\delta(E_j-E^*)\nonumber\\
&\times&(\sin\theta_j)^{d-4}{\rm d}(\cos\theta_j)\;(\sin\psi_j)^{d-5}{\rm d}(\cos\psi_j){\rm d}\Omega_{d-4}\,.
\end{eqnarray} 
with 
\beq
E^*=\frac{1}{2E_{cm}}(E_{cm}^2-m^2),
\eeq
where $E_{cm}$ denotes the centre-of-mass energy of the two-particle system. For all soft-factor contributions, the delta function in the phase-space measure constrains the unresolved-gluon energy 
to $E_j=E^*$. Consequently, the energy integration can be performed trivially, leaving only the angular integrations to be performed.

With the massive antenna phase-space now fully determined, we proceed with the analytical integration of the massive soft factor defined in~\eqref{eq:SF0m}. 
The integrated soft-factor normalised appropriately is defined by
\begin{equation}
{\cal S}^{0m}_{0m}(s_{ac},s_{IK})=\frac{1}{C(\epsilon)}\int {\rm d}X_{i,jk}(p_j,p_k;p_i,q) S_{ajc}^{0m}.
\end{equation}

To integrate the massive soft factor, we choose a reference frame in which the massive spectator $p_c$ defines the $z$-axis, while the massless spectator $p_a$ lies in the $xz$-plane.
We note that, while the integration is performed over the antenna phase space defined by the hard radiators $p_i$, $p_k$, 
the hard momenta $p_a$, $p_c$, need not coincide with
$p_i$ and $p_k$. They may instead correspond to any pair of on-shell hard momenta. The corresponding four-momenta are,
\begin{eqnarray}
p_c&=&E_c(1,0,0,\beta)\nonumber\\
p_a&=&E_a(1,\sin\theta_a,0,\cos\theta_a)\nonumber\\
p_j&=&E_j(1,\sin\theta_j\cos\psi_j,\sin\theta_j\sin\psi_j,\cos\theta_j)\nonumber
\end{eqnarray}
with $\beta^2=1-\frac{m^2}{E_c^2}$.
For later convenience, we also introduce the mapped radiator momenta $p_I$ and $p_K$ in the reduced initial-final antenna configuration, obtained after clustering the unresolved gluon~\cite{Abelof:2011ap},
\begin{eqnarray}
p_I&=&x_i p_i\nonumber\\
p_K&=&p_j+p_k-(1-x_i)p_i\,.
\end{eqnarray}
These enter only through the invariants $s_{cI}=2p_c\cdot p_I$ and $s_{cK}=2p_c\cdot p_K$, which are used below to express the velocity of the massive spectator.

Since the soft factor naturally decomposes into a mass-dependent contribution and a mass-independent contribution, 
\begin{equation}
S_{ajc}=\frac{2s_{ac}}{s_{aj}s_{jc}}-\frac{2m^2}{s_{jc}^2}=S_{ajc}^{{\rm mi}}+S_{ajc}^{{\rm md}}
\end{equation}
it is convenient to evaluate these two terms separately. 
We begin by considering the mass-dependent term proportional to $-2m^2/s_{jc}^2$. The mass-dependent soft-factor integrated over the initial-final massive antenna phase space reads,
\begin{eqnarray}
{\cal S}_{0m}^{0m;md}(s_{ac},s_{IK})&=&-\int \frac{8\pi^2 e^{\epsilon\gamma_e}}{(4\pi)^{\epsilon}}\left(\frac{Q^2+m^2}{2\pi}\right)\,\frac{(2\pi)^{2-d}m^2}{8E_c^2E_{cm}}\,E_j^{d-5}{\rm d}E_j\delta(E_j-E^*)\nonumber\\
&\times&\frac{(\sin\theta_j)^{d-4}{\rm d}(\cos\theta_j)\;(\sin\psi_j)^{d-5}{\rm d}(\cos\psi_j)}{(1-\beta\cos\theta_j)^2}{\rm d}\Omega_{d-4}
\end{eqnarray}

Since the integration over the gluon energy is fixed by the energy-conservation delta function, the remaining integration is purely angular and yields,
\begin{equation}
\frac{2\pi^{3/2-\epsilon}}{\Gamma\left(\frac{3}{2}-\epsilon\right)}\;{}_2F_1\Big(1,\frac{3}{2}, \frac{3}{2} - \epsilon, \beta^2\Big).
\end{equation}
Since the final state momenta $p_j$ and $p_k$ are integrated over, the only kinematical variables that the integrated soft-factor can depend on 
are $\beta$ and the invariants constructed from the incoming momenta $q$ and $p_i$, that is $Q^2$ and $p_i \cdot q$. 
It is convenient to express this dependence in terms of the dimensionless variables,
\begin{equation}
x_0=\frac{Q^2}{Q^2+m^2}
\end{equation}
and $x_i$, the momentum fraction appearing in the factorization of the initial-final massive antenna phase space~\cite{Abelof:2011jv},
\begin{equation}
x_i=\frac{Q^2+m^2}{2p_i \cdot q}.
\end{equation}
If the initial state radiator is 1, $(p_i=p_1)$ the center of mass energy is given by
\begin{equation}
E_{cm}^2=(q+p_1)^2=\frac{Q^2(1-x_1)+m^2}{x_1},
\end{equation}
and the final result for the mass-dependent integrated soft factor expressed in terms of $\beta, x_0, x_1$ reads,
\begin{eqnarray}
{\cal S}_{0m}^{0m;md}(s_{ac},s_{IK})&=&-\frac{e^{\epsilon\gamma_e}\sqrt{\pi}}{\Gamma\left(\frac{3}{2}-\epsilon\right)}
2^{-1+2\epsilon}\left(Q^2\right)^{-\epsilon}x_0^{\epsilon}\;(1-x_1)^{-1-2\epsilon}\;x_1^{1+\epsilon}(1-x_0x_1)^{\epsilon}\nonumber\\
&\times&(1-\beta^2)\;{}_2F_1\Big(1,\frac{3}{2}, \frac{3}{2} - \epsilon, \beta^2\Big),
\label{eq:mdSFIF0m}
\end{eqnarray}
with
\begin{eqnarray}
\beta^2&=&1-\frac{4 m^2 x_1 ( Q^2(1-x_1)+m^2)}{(s_{cI}(1-x_1)+s_{cK}x_1)^2}\,.
\end{eqnarray}

To obtain the integrated soft factor in expanded form, two separate steps are required. First, the hypergeometric function 
and the remaining regular part of the integrand are expanded around $\epsilon=0$. Second, the endpoint singular factor 
in Eq.~\eqref{eq:mdSFIF0m} has to be expanded in distributions. 
In particular, we use,
\begin{equation}
(1-x_1)^{-1-n\epsilon}=-\frac{1}{n\epsilon}\delta(1-x_{1})+\sum_{m=0}^{\infty}\frac{(-n\epsilon)^m}{m!}{\cal D}_{m}(x_1)
\end{equation}
where
\begin{equation}
{\cal D}_{m}(x_1)=\left(\frac{\log^{m}(1-x_1)}{1-x_1}\right)_+\;
\end{equation}
denotes the standard plus distribution. 

The Laurent expansion isolates the explicit endpoint pole at $x_1=1$ and allows the mass dependent integrated soft factor to be written as a Laurent series in $\epsilon$,
\begin{eqnarray}
{\cal S}_{0m}^{0m;md}(s_{ac},s_{IK})&=&\frac{1}{2\epsilon}\delta(1-x_1)-{\cal D}_0(x_1)+1\nonumber\\
 &+&\delta(1-x_1)\Bigg[ \frac{1}{2\beta}\left(\log\left(\frac{1+\beta}{1-\beta}\right)+\beta\log(1-x_0)\right) \Bigg].
\end{eqnarray}
For simplicity, we omit the common overall factor $(Q^2)^{-\epsilon}$
arising from the dimensional regularisation of the phase-space measure in all Laurent expansions. 

We now turn to the mass-independent contribution to the integrated soft factor in~\eqref{eq:SF0m}. Its contribution integrated over the the initial-final phase space reads,
\begin{eqnarray}
{\cal S}_{0m}^{0m;mi}(s_{ac},s_{IK})&=&\int \frac{8\pi^2 e^{\epsilon\gamma_e}}{(4\pi)^{\epsilon}}\left(\frac{Q^2+m^2}{2\pi}\right)\,\frac{(2\pi)^{2-d}s_{ac}}{8E_a E_c E_{cm}}\,E_j^{d-5}{\rm d}E_j\delta(E_j-E^*)\nonumber\\
&\times&\frac{(\sin\theta_j)^{d-4}{\rm d}(\cos\theta_j)\;(\sin\psi_j)^{d-5}{\rm d}(\cos\psi_j)}{(1-\beta\cos\theta_j)(1-\sin\theta_a\sin\theta_j\sin\psi_j-\cos\theta_a\cos\theta_j)}{\rm d}\Omega_{d-4}\,.
\end{eqnarray}
The integration over the gluon energy is exactly identical to that performed for the mass-dependent contribution and 
is fixed by the energy-conservation delta function. The remaining angular integration is, however, 
more involved, and can be expressed in terms of the Appell function of the first kind~\cite{Somogyi:2011ir},
\begin{eqnarray}
(4\pi)^{1-\epsilon}\frac{2}{1-\beta\cos\theta_a}\frac{\Gamma(-\epsilon)}{\Gamma(1-2\epsilon)}
F_1\Big(1, -\epsilon, -\epsilon, 1 - 2 \epsilon, -\frac{\beta(1+\cos\theta_a)}{1-\beta\cos\theta_a}, \frac{\beta(1-\cos\theta_a)}{1-\beta\cos\theta_a}\Big)\hspace{0.6cm}
\end{eqnarray}
For the present calculation, the Appell function is expanded for small $\epsilon$ according to
\begin{eqnarray}
F_1\Big(1, -\epsilon, -\epsilon, 1 - 2 \epsilon, x, y\Big)&=&1+\epsilon\log\left((1-x)(1-y)\right)\nonumber\\
&+&\epsilon^2\left(-\frac{1}{2}\log\left(\frac{1-x}{1-y}\right)^2-2\textrm{Li}_2(x)-2\textrm{Li}_2(y)\right).
\end{eqnarray}

Since in our momentum parametrisation $s_{ac}=2E_aE_c(1-\beta\cos\theta_a)$ it is convenient to introduce the variable,
\begin{equation}
x_{ac;IK}=\frac{s_{ac}}{4E_aE_c}=\frac{s_{ac}(Q^2(1-x_1)+m^2)}{(s_{Ia}(1-x_1)+x_1s_{aK})(s_{Ic}(1-x_1)+x_1s_{cK})}
\end{equation}
and the final result for the mass-independent integrated soft factor expressed in terms of $\beta, x_{ac;IK}, x_0, x_1$ reads,
\begin{eqnarray}
{\cal S}_{0m}^{0m;md}(s_{ac},s_{IK})&=&\frac{e^{\epsilon\gamma_e}\Gamma(-\epsilon)}{\Gamma(1-2\epsilon)}\left(Q^2\right)^{-\epsilon} x_0^{\epsilon}
x_1^{1+\epsilon}(1-x_1)^{-1-2\epsilon}(1-x_0x_1)^{\epsilon}\nonumber\\
&\times&F_1\Big(1, -\epsilon, -\epsilon, 1 - 2 \epsilon, 1-\frac{1+\beta}{2x_{ac;IK}}, 1-\frac{1-\beta}{2x_{ac;IK}} \Big)\,.
\end{eqnarray}
Following the same procedure as for the mass-dependent contribution, we expand the result around $\epsilon=0$. 
Also here, the endpoint is expanded in distributions, yielding the Laurent expansion of the mass-independent integrated soft factor,
\begin{eqnarray}
{\cal S}_{0m}^{0m;md}(s_{ac},s_{IK})&=&
\frac{1}{2\epsilon^2}\delta(1-x_1)\nonumber\\
&+&\frac{1}{\epsilon}\delta(1-x_1)\;\frac{1}{2}\big(-2\log(2)+\log(1-x_0)+\log(1-\beta^2)-2\log(x_{ac;ik})\big)\nonumber\\
&+&\frac{1}{\epsilon}\big(
 1-{\cal D}_0(x_1)\big)\nonumber\\
 &+&\delta(1-x_1)\Bigg(-\log(2)\log(1-x_0)+\frac{1}{4}\log(1-x_0)^2-\frac{1}{4}\log\left(\frac{1+\beta}{1-\beta}\right)^2\nonumber\\
 &&\hspace{0.4cm}+\frac{1}{2}\log(1-x_0)\log(1-\beta^2)-\log(1-x_0)\log(x_{ac;ik}) \nonumber\\
 &&\hspace{0.4cm}
 - \textrm{Li}_2\Big( 1 - \frac{1 - \beta}{2x_{ac;ik}} \Big)
 -\textrm{Li}_2\Big( 1 - \frac{1 + \beta}{2x_{ac;ik}} \Big)
 -\frac{\pi^2}{8}
 \Bigg)\nonumber\\
&+&{\cal D}_0(x_1)\Big(2\log(2)-\log(1-x_0)-\log(1-\beta^2)+2\log(x_{ac;ik})\Big)+2{\cal D}_1(x_1)\nonumber\\
&-&2\log(2)+\frac{\log(1-x_0)}{1-x_1}-2\log(1-x_1)+\log(x_1)-\frac{\log(x_1)}{1-x_1}\nonumber\\
&+&\log(1-x_0x_1)-\frac{\log(1-x_0x_1)}{1-x_1}
+\frac{\log(1-\beta^2|_{x_1=1})}{1-x_1}+\log(1-\beta^2)
\nonumber\\
&-&\frac{\log(1-\beta^2)}{1-x_{1}}-\frac{2\log(x_{ac;ik}|_{x_1=1})}{1-x_1}-2\log(x_{ac;ik})
+\frac{2\log(x_{ac;ik})}{1-x_1}\,.
\end{eqnarray}

\subsection{Massless soft factor integrated over the initial-final massive  phase space}
\label{sec:SFint2}
We now turn to the integration of the massless soft factor, also present in unintegrated form in Eq.~\eqref{eq:qgsnnlo}.
Although this soft factor involves two massless radiators, we integrate it over the same initial-final massive antenna phase space as in the previous subsection. This choice is motivated by the structure of the subtraction terms given in Eq.~\eqref{eq:qgsnnlo}, where massless and massive large-angle soft contributions appear together.

In this case, for two massless spectators $p_a^2=p_c^2=0$, the relevant soft-factor is the massless eikonal factor,
\begin{equation}
S_{ajc}=\frac{2s_{ac}}{s_{aj}s_{jc}},
\end{equation}
with $(s_{rs}=2p_r\cdot p_s)$. 
The integrated soft factor normalised appropriately is defined by
\begin{equation}
{\cal S}_{00}^{0m}(s_{ac},s_{IK})
=
\frac{1}{C(\epsilon)}
\int {\rm d}X_{i,jk}(p_j,p_k;p_i,q)\,
S_{ajc},
\label{eq:int_massless_soft_0m}
\end{equation}
where
\begin{equation}
C(\epsilon)
=
(4\pi)^\epsilon
\frac{-e^{\epsilon\gamma_E}}{8\pi^2}.
\end{equation}

We perform the integration in the following reference frame,
\begin{eqnarray}
p_a&=&E_a(1,0,0,1)\nonumber\\
p_c&=&E_c(1,\sin\theta_c,0,\cos\theta_c)\nonumber\\
p_j&=&E_j(1,\sin\theta_j\cos\psi_j,\sin\theta_j\sin\psi_j,\cos\theta_j).
\nonumber
\end{eqnarray}

Using this parametrization in the phase-space integral, the massless integrated soft-factor contribution can be written as
\begin{eqnarray}
{\cal S}_{00}^{0m}(s_{ac},s_{IK})&=&\int \frac{8\pi^2 e^{\epsilon\gamma_e}}{(4\pi)^{\epsilon}}\left(\frac{Q^2+m^2}{2\pi}\right)\,\frac{(2\pi)^{2-d}s_{ac}}{8E_a E_c E_{cm}}\,E_j^{d-5}{\rm d}E_j\delta(E_j-E^*)\nonumber\\
&\times&\frac{(\sin\theta_j)^{d-4}{\rm d}(\cos\theta_j)\;(\sin\psi_j)^{d-5}{\rm d}(\cos\psi_j)}{(1-\cos\theta_j)(1-\sin\theta_c\sin\theta_j\sin\psi_j-\cos\theta_c\cos\theta_j)}{\rm d}\Omega_{d-4}\,.
\label{eq:00PS0mSFIF}
\end{eqnarray}
The energy integration proceeds exactly as in the previous cases. The remaining angular integration yields,
\begin{equation}
-(2\pi)\frac{(4\pi)^{\epsilon}\,\Gamma(1-\epsilon)}{\Gamma(1-2\epsilon)}\left(\frac{1}{\epsilon}\right){}_2F_1\left(1,1,1-\epsilon,\cos^2\left(\frac{\theta_c}{2}\right)\right).
\end{equation}
Since in our parametrization $s_{ac}=2E_aE_c(1-\cos\theta_c)$ it is convenient to introduce the variable
\begin{eqnarray}
x_{ac;IK}&=&\frac{s_{ac}}{4E_aE_c}=\frac{s_{ac}(Q^2(1-x_1)+m^2)}{(s_{Ia}(1-x_1)+x_1s_{aK})(s_{Ic}(1-x_1)+x_1s_{cK})}\,,
\end{eqnarray}
and the final result for the massless integrated soft factor expressed in terms of $x_{ac;IK}, x_0, x_1$ reads,
\begin{eqnarray}
{\cal S}_{00}^{0m}(s_{ac},s_{IK})&=&\frac{e^{\epsilon\gamma_e}\Gamma(1-\epsilon)}{\Gamma(1-2\epsilon)}\left(Q^2\right)^{-\epsilon}x_0^{\epsilon}x_1^{1+\epsilon}(1-x_1)^{-1-2\epsilon}
(1-x_0x_1)^{\epsilon}\nonumber\\
&\times&\left(-\frac{2}{\epsilon}\right)x_{ac;ik}\,{}_2F_1\left(1,1,1-\epsilon,1-x_{ac;ik})\right).
\label{eq:00PS0mSFIFV2}
\end{eqnarray}

After expanding the result around $\epsilon=0$ and expressing the endpoint singularity in terms of distributions, the Laurent expansion of the massless integrated soft factor is obtained. It reads 
\begin{eqnarray}
{\cal S}_{00}^{0m}(s_{ac},s_{IK})&=& \frac{1}{\epsilon^2}\delta(1-x_1)+
\frac{1}{\epsilon}\Big[\delta(1-x_1)\;\big(\log(1-x_0)-\log(x_{ac;ik})\big)-2{\cal D}_0(x_1)+2\Big]\nonumber\\
&+&\delta(1-x_1)\Bigg(\frac{1}{2}\log(1-x_0)^2-\log(1-x_0)\log(x_{ac;ik})-\textrm{Li}_2\left(-\frac{1-x_{ac;ik}}{x_{ac;ik}}\right)\nonumber\\&-&\frac{\pi^2}{4}\Bigg)
+{\cal D}_0(x_1)\Big(-2\log(1-x_0)+2\log(x_{ac;ik})\Big)+4{\cal D}_1(x_1)\nonumber\\
&+&\frac{2\log(1-x_0)}{1-x_1}-4\log(1-x_1)
+2\log(x_1)-\frac{2\log(x_1)}{1-x_1}+2\log(1-x_0x_1)\nonumber\\
&-&\frac{2\log(1-x_0x_1)}{1-x_1}-\frac{2\log(x_{ac;ik}|_{x_1=1})}{1-x_1}
-2\log(x_{ac;ik})+\frac{2\log(x_{ac;ik})}{1-x_1}\,.
\label{eq:00PS0mSFIFv3}
\end{eqnarray}

A useful and non-trivial consistency check of the above result is provided by taking its massless limit. Since the dependence on the heavy-quark mass enters only through the 
initial-final massive antenna phase space, while the soft factor itself is independent of the mass, the limit $m\to0$ should reproduce the known expression for the 
massless integrated soft factor over the initial-final massless antenna phase space presented in Ref.~\cite{Daleo:2009yj}.
To perform this comparison, both results must first be expanded in distributions. 
The massless limit is then taken in the Laurent expansion derived above and found in agreement with the corresponding expansion of Eq.~(2.22) of Ref.~\cite{Daleo:2009yj}.

\subsection{Numerical validation of the integrated large-angle soft subtraction}
\label{sec:SFnumcheck}
In this section we validate the implementation of the integrated soft factors derived above by means of a numerical consistency check within the antenna subtraction framework. The validation is performed by introducing an auxiliary test contribution modelled on the large-angle soft subtraction term appearing in Eq.~\eqref{eq:qgsnnlo}. It combines the large-angle soft factors with suitable three-parton antenna functions multiplied by reduced Born matrix elements in such a way that the complete integrand is infrared finite. The corresponding analytically integrated contribution is then added at the real-virtual level. Since the same quantity is added and subtracted, the double-real and real-virtual contributions must integrate to identical results with opposite sign.

The comparison is carried out as a full Monte Carlo integration over the $t\bar{t}$ production phase space, thereby providing a stringent test of both the analytical integration and its implementation.
The test contribution at real-real level is given by
\begin{eqnarray}
{\rm d}\hat{\sigma}_{NNLO}^{\textrm{test},RR}&=&
\Big(
S_{\bar{1}62}
-S_{\bar{\bar{1}}\tilde6\bar2}
-S_{\bar16(36)}
+S_{\bar{\bar1}\tilde6\widetilde{(36)}}
-S_{26(36)}
+S_{\bar2\tilde6\widetilde{(36)}}
\Big)
\nonumber\\
&\times&
G_3^0(5_g,\hat1_q,\hat2_{\bar q})\,
C_0^0(\hat{\bar{\bar1}}_q,\widetilde{(36)}_Q,\tilde4_Q,\hat{\bar2}_q)
\nonumber\\
&-&
A_3^0(\hat1_q,6_g,\hat2_q)\,
G_3^0(\tilde5_g,\hat{\bar1}_q,\hat{\bar2}_q)\,
C_0^0(\hat{\bar{\bar1}}_q,\tilde3_Q,\tilde4_Q,\hat{\bar{\bar2}}_q)
\nonumber\\
&+&
G_3^0(5_g,\hat1_q,\hat2_q)\,
A_3^0(\hat{\bar1}_q,\tilde6_g,\hat{\bar2}_q)\,
C_0^0(\hat{\bar{\bar1}}_q,\tilde3_Q,\tilde4_Q,\hat{\bar{\bar2}}_q)
\nonumber\\
&+&
A_3^0(\hat1_q,6_g,3_Q)\,
G_3^0(5_g,\hat{\bar1}_q,\hat2_q)\,
C_0^0(\hat{\bar{\bar1}}_q,\widetilde{(36)}_Q,\tilde4_Q,\hat{\bar2}_q)
\nonumber\\
&-&
G_3^0(5_g,\hat1_q,\hat2_q)\,
A_3^0(\hat{\bar1}_q,\tilde6_g,\widetilde3_Q)\,
C_0^0(\hat{\bar{\bar1}}_q,\widetilde{(36)}_Q,\tilde4_Q,\hat{\bar2}_q)
\nonumber\\
&+&
A_3^0(\hat2_q,6_g,3_Q)\,
G_3^0(5_g,\hat1_q,\hat{\bar2}_q)\,
C_0^0(\hat{\bar1}_q,\widetilde{(36)}_Q,\tilde4_Q,\hat{\bar{\bar2}}_q)
\nonumber\\
&-&
G_3^0(5_g,\hat1_q,\hat2_q)\,
A_3^0(\hat{\bar2}_q,\tilde6_g,\widetilde3_Q)\,
C_0^0(\hat{\bar1}_q,\widetilde{(36)}_Q,\tilde4_Q,\hat{\bar{\bar2}}_q).
\label{eq:RRsfIFtest}
\end{eqnarray}
In Eq.~\eqref{eq:RRsfIFtest}, the momenta $p_1$ and $p_2$ denote the initial-state massless partons, while $p_3$ and $p_4$ are the final-state massive quarks. 
The momenta $p_5$ and $p_6$ correspond to final-state massless partons. The large-angle soft subtraction terms are first evaluated using an initial-final antenna mapping~\cite{Abelof:2011ap} in which the momenta
($p_1,p_3,p_6$) are mapped onto the reduced configuration ($p_{\bar{1}},p_{\widetilde{36}}$). Subsequently, an initial-initial 
mapping~\cite{Abelof:2011ap} is applied to construct the Born-level kinematics entering the reduced matrix elements. 
This second mapping redefines the initial-state momenta $p_{\bar{1}}$ and $p_2$ into the remapped momenta $p_{\bar{\bar{1}}}$ and $p_{\bar{2}}$, which differ from the original momenta only by a momentum-fraction shift and are denoted by hats. The final-state momenta are boosted accordingly as necessary to satisfy overall momentum conservation, and are denoted with tildes. 

Using the above kinematics, the test contribution in Eq.~\eqref{eq:RRsfIFtest} consists of a combination of massless and one-mass three-parton $X_3^0$ antennae, whose integrated forms have been derived 
and validated in previous works~\cite{Abelof:2011jv,Daleo:2006xa}, together with the new large-angle massive soft factors derived in Sections~\ref{sec:SFint1},~\ref{sec:SFint2}. Although the individual terms in~\eqref{eq:RRsfIFtest} contain soft and collinear singularities, 
these singularities cancel in the complete combination. Consequently, the resulting integrand is infrared finite over the entire phase space and can be integrated directly in four dimensions using Monte Carlo methods.

At the real-virtual level we add back
\begin{eqnarray}
{\rm d}\hat{\sigma}_{NNLO}^{\textrm{test},RV}&=&
-G_3^0(5_g,\hat{1}_q,\hat{2}_q)\Big(
{\cal S}_{00}^{0m}(s_{12},s_{13})
-\tilde{{\cal S}}_{00}^{0m}(s_{\bar{1}\bar{2}},s_{\bar{1}\tilde{3}})
-{\cal S}_{0m}^{0m}(s_{13},s_{13})
+\tilde{{\cal S}}_{0m}^{0m}(s_{\bar{1}\tilde{3}},s_{\bar{1}\tilde{3}})
\nonumber\\
&&\hspace{0.5cm}-{\cal S}_{0m}^{0m}(s_{23},s_{13})
+\tilde{{\cal S}}_{0m}^{0m}(s_{\bar{2}\tilde{3}},s_{\bar{1}\tilde{3}})
-{\cal A}_{3,qq}^0(s_{12})+{\cal A}_{3,q}^{0m}(s_{13})
+{\cal A}_{3,q}^{0m}(s_{23})
\nonumber\\
&&\hspace{0.5cm}+{\cal A}_{3,qq}^0(s_{\bar{1}\bar{2}})-{\cal A}_{3,q}^{0m}(s_{\bar{1}\tilde{3}})-{\cal A}_{3,q}^{0m}(s_{\bar{2}\tilde{3}})
\Big)C_0^{0}(\hat{\bar{1}}_q,\tilde{3}_Q,\tilde{4}_Q,\hat{\bar{2}}_q)\nonumber\\
&-&A_3^0(\hat{1}_q,5_g,\hat{2}_q)\;{\cal G}_{3,qq}^0(s_{12})\;C_0^{0}(\hat{\bar{1}}_q,\tilde{3}_Q,\tilde{4}_Q,\hat{\bar{2}}_q)\nonumber\\
&+&A_3^0(\hat{1}_q,5_g,3_Q)\;{\cal G}_{3,qq}^0(s_{12})\;C_0^{0}(\hat{\bar{1}}_q,\tilde{3}_Q,\tilde{4}_Q,\hat{2}_q)\nonumber\\
&+&A_3^0(\hat{2}_q,5_g,3_Q)\;{\cal G}_{3,qq}^0(s_{12})\;C_0^{0}(\hat{1}_q,\tilde{3}_Q,\tilde{4}_Q,\hat{\bar{2}}_q)\nonumber\\
&&\nonumber\\
&+&A_3^0(\hat{1}_q,5_g,\hat{2}_q)\;{\cal G}_{3,qq}^0(s_{\bar{1}\bar{2}})\;C_0^{0}(\hat{\bar{1}}_q,\tilde{3}_Q,\tilde{4}_Q,\hat{\bar{2}}_q)\nonumber\\
&-&A_3^0(\hat{1}_q,5_g,3_Q)\;{\cal G}_{3,qq}^0(s_{\bar{1}2})\;C_0^{0}(\hat{\bar{1}}_q,\tilde{3}_Q,\tilde{4}_Q,\hat{2}_q)\nonumber\\
&-&A_3^0(\hat{2}_q,5_g,3_Q)\;{\cal G}_{3,qq}^0(s_{1\bar{2}})\;C_0^{0}(\hat{1}_q,\tilde{3}_Q,\tilde{4}_Q,\hat{\bar{2}}_q).
\label{eq:RVsfIFtest}
\end{eqnarray}
As can be seen from Eqs.~\eqref{eq:RRsfIFtest} and \eqref{eq:RVsfIFtest}, the large-angle soft subtraction terms involve both massless and massive spectator configurations, 
making this test simultaneously sensitive to the massless and one-mass soft factors integrated over the initial-final massive 
antenna phase space.
Consequently, agreement between the integrated double-real and real-virtual contributions provides a direct validation of both integrated soft-factor contributions.

As a first validation, we verified analytically that all explicit infrared poles cancel in the real-virtual contribution, i.e in Eq.~\eqref{eq:RVsfIFtest}. 
This provides a stringent check of the Laurent expansions of the integrated soft factors derived in Sections~\ref{sec:SFint1},~\ref{sec:SFint2}, since the cancellation requires the pole 
structure of the newly derived integrals to match exactly that of the remaining integrated antenna functions.
The remaining finite contributions are then integrated numerically over the full $t\bar{t}$ production phase space using NNLOJET yielding,
\begin{eqnarray}
{\rm d}\sigma^{RR}&=&\phantom{-}117.90 \pm 0.50,\nonumber\\
{\rm d}\sigma^{RV}&=&-117.89 \pm 0.07.\nonumber
\end{eqnarray}
The two contributions are equal in magnitude and opposite in sign within the Monte Carlo uncertainties, demonstrating the correct implementation of the integrated massless and one-mass soft factors derived in this work.
\section{Numerical results} 
\label{sec:nnlonumerics}
In this section we present numerical results obtained with our implementation in the NNLOJET parton-level generator of the NNLO contributions to heavy-quark pair production at hadron colliders in the partonic channels
$ab\to t\bar t+X$, with
$ab=qg(\bar q g)$,
$qq(\bar q\bar q)$,
$qq'(\bar q\bar q')$,
where $q$ and $q'$ denote massless quarks of different flavour. All colour structures are included.
The calculation is performed within the massive antenna subtraction framework as
described in the previous sections and incorporates, in particular, the integrated initial-final soft factors and the massive antenna convolutions derived in this work for the first time.

For the numerical results presented below, we consider proton-proton collisions at a centre-of-mass energy of $\sqrt{s}=13$ TeV. Throughout, we employ the NNPDF4.0 NNLO parton distribution functions. The strong coupling is evolved at three-loop order with $\alpha_s(m_Z)=0.118$, and the pole mass of the top quark is fixed to $m_t=173.3$ GeV. In all processes, the 
phase-space generation imposes a technical cut $y_0=10^{-6}$, requiring
all Mandelstam invariants to remain above $y_0\hat{s}$. This cut is sufficiently small compared to the physical scales of the problem so as not to affect the numerical results.

The NNLO coefficients ($\delta\sigma_{NNLO}$) entering the inclusive heavy-quark pair production cross section in the partonic channels listed above are compared with the corresponding predictions from {\tt Top++}~\cite{Czakon:2011xx,Czakon:2012zr,Czakon:2012pz} The results are presented for the central renormalization and factorisation scale choice, $\mu_R=\mu_F=m_t$ together with the two factorization-scale variations ($\mu_R,\mu_F=m_t,m_t/2$) and ($m_t,2m_t$). Throughout the following, the subscript $qg$ denotes the combined contribution from the channels $qg$ and $\bar q g$, while the subscript $q(\bar{q})q'$, denote the combined contributions from the charge-conjugate channel pairs $qq(\bar q\bar q)$, $qq'(\bar q\bar q')$, respectively, where $q$ and $q'$ denote quarks of different flavour.
The corresponding results are collected 
in Table~\ref{tab:NNLOcoefficientsQQ} and Table~\ref{tab:NNLOcoefficientsQG}.

The agreement between NNLOJET and {\tt Top++} is excellent for both partonic channels. Although only a representative subset of scale choices is shown, the comparison was verified for the complete seven-point scale variation.

\begin{table}[htb!]
\centering
\begin{tabular}{lcc}
\hline
$\delta\sigma_{NNLO,q(\bar{q})q'}(\mu_R,\mu_F)$ [pb] & NNLOJET & {\tt Top++} \\
\hline
$(m_t,m_t)$         & $0.4788(5)$  & $0.4773$ \\
$(m_t,m_t/2)$       & $1.8133(9)$  & $1.8134$ \\
$(m_t,2m_t)$        & $0.6996(4)$  & $0.6973$ \\
\hline
\end{tabular}
\caption{NNLO coefficient in picobarn [pb] for the $q(\bar{q})q'$ channel compared to the corresponding prediction from {\tt Top++}
at $\sqrt{s}=13$ TeV. The quoted NNLOJET
uncertainty denoted in parenthesis on the last digit corresponds to the Monte Carlo integration error.}
\label{tab:NNLOcoefficientsQQ}
\end{table}

\begin{table}[t]
\centering
\begin{tabular}{lcc}
\hline
$\delta\sigma_{NNLO,qg}(\mu_R,\mu_F)$ [pb] & NNLOJET & {\tt Top++} \\
\hline
$(m_t,m_t)$         & $-2.06(2)$            & $-2.056$ \\
$(m_t,m_t/2)$       & $\phantom{-}8.28(2)$ & $\phantom{-}8.301$ \\
$(m_t,2m_t)$        & $-15.40(3)$          & $-15.387$ \\
\hline
\end{tabular}
\caption{NNLO coefficient in picobarn [pb] for the $qg$ channel compared to the corresponding prediction from {\tt Top++}
at $\sqrt{s}=13$ TeV. The quoted NNLOJET
uncertainty denoted in parenthesis on the last digit corresponds to the Monte Carlo integration error.}
\label{tab:NNLOcoefficientsQG}
\end{table}

For the $q(\bar q)q'$ contribution, the relative difference between the two results remains below $0.4\%$ for all scale choices. In the $qg$ channel, the NNLOJET and \texttt{Top++} central values agree very well, with relative differences below 0.3\% for all scale choices and within the quoted NNLOJET Monte Carlo integration uncertainties, which are at the percent level for this channel.
The {\tt Top++} reference values were obtained using the {\tt Precision=3} setting. No numerical uncertainty is returned by the program for the corresponding predictions. The uncertainties quoted in Tables~\ref{tab:NNLOcoefficientsQQ} and \ref{tab:NNLOcoefficientsQG} therefore correspond exclusively to the Monte Carlo integration errors of the NNLOJET calculation and are indicated in parentheses on the last quoted digit. Given the range of variation exhibited by the NNLO coefficients under changes of the renormalization and factorization scales, the observed agreement provides a stringent validation of the implementation of the NNLO corrections in NNLOJET. The results confirm the correctness of the complete subtraction framework, including the integrated initial-final soft factors and massive antenna convolutions derived in this work.

\section{Conclusions}
\label{sec:conclusions}
In this paper, we have completed the antenna-subtraction treatment of the $qq$, $qq'$ and $qg$ partonic channels contributing to heavy-quark pair production at ${\cal O}(\alpha_s^4)$ in QCD and implemented them within the NNLOJET~\cite{NNLOJET:2025rno} framework. Building upon the previously available double-real subtraction terms for the purely fermionic channels, we derived the subtraction contributions entering the real-virtual and double-virtual levels, thereby completing the NNLO description of the $qq$ and $qq'$ channels in full colour. In addition, we presented the complete computation up to ${\cal O}(\alpha_s^4)$ of the $qg$-channel contribution to heavy-quark pair production in full colour. The infrared structure of all three channels was analysed systematically and compact representations of the subtraction terms at the virtual-virtual level in terms of massless and massive integrated dipole operators were obtained.  

The completion of the cross section computation including contributions at ${\cal O}(\alpha_s^4)$ required several new analytic ingredients:  
 we derived the integrated initial-final massive and massless soft factors over the massive antenna initial-final phase space. These results were validated through dedicated numerical checks, providing a direct test of both, the newly derived analytic expressions for the massive soft factors and their correct implementation in the NNLOJET parton-level generator.

In addition, we evaluated, for the first time, convolution integrals involving integrated massive three-parton antennae with integrated massless antennae and one-loop mass-factorisation kernels. Together with the newly integrated soft factors, these results extend the integrated antenna library for processes involving massive fermions and constitute important ingredients for future NNLO calculations using the massive antenna formalism, beyond the specific hadronic top pair production process considered here.

The correctness of the antenna subtraction framework was further established through several independent numerical checks. We verified the local behaviour of the subtraction terms in all unresolved limits at real-real and real-virtual-level and demonstrated the complete analytic cancellation of the explicit infrared poles after combining the integrated antenna contributions with the corresponding mass-factorisation counterterms at real-virtual and virtual-virtual levels. 

Finally, the complete NNLO implementation within NNLOJET, combining the full-colour matrix elements from OpenLoops with all newly derived subtraction terms, was benchmarked against the inclusive predictions of {\tt Top++}, finding excellent agreement.

The results presented here provide the complete NNLO treatment of the $qq$, $qq'$ and $qg$ partonic channels within the antenna-subtraction formalism. Together, these channels represent an essential component of a fully differential NNLO description of heavy-quark pair production. Their implementation within NNLOJET enables precision phenomenological studies with arbitrary fiducial selections and fully differential predictions based on the complete kinematic information of the final state. The remaining partonic channels required for a complete and full colour NNLO description of top-quark pair production arise from the $gg$ and $q\bar{q}$-channels. Their treatment will require a further extension of the integrated antenna library and is left for future work.

\begin{acknowledgments}
 AG is supported by the Swiss National Science Foundation (SNF) under contract 200021-231259. JP acknowledges financial support from the Portuguese Funda\c{c}\~{a}o para a Ci\^encia e Tecnologia (FCT) under contract 2024.11719.CEECIND.
\end{acknowledgments}

\bibliographystyle{JHEP}
\bibliography{bib_tt_qg}
            
\end{document}